\documentclass[alpha-refs]{wiley-article}

\usepackage{graphicx}
\usepackage[space]{grffile}
\usepackage{latexsym}
\usepackage{textcomp}
\usepackage{longtable}
\usepackage{tabulary}
\usepackage{booktabs,array,multirow}
\usepackage{amsfonts,amsmath,amssymb}
\usepackage{natbib}
\usepackage{url}
\usepackage{hyperref}
\hypersetup{colorlinks=false,pdfborder={0 0 0}}
\usepackage{etoolbox}
\newif\iflatexml\latexmlfalse

\AtBeginDocument{\DeclareGraphicsExtensions{.pdf,.PDF,.eps,.EPS,.png,.PNG,.tif,.TIF,.jpg,.JPG,.jpeg,.JPEG}}

\usepackage[utf8]{inputenc}
\usepackage[english]{babel}

\usepackage{siunitx}
\usepackage{overpic}
\usepackage{caption}
\usepackage{amsfonts}
\usepackage{authblk}
\usepackage{orcidlink}
\usepackage{subcaption}
\usepackage{makecell}
\usepackage{colortbl}

\iflatexml

\else
\fi

\usepackage{lineno}

\corraddress{Fabian Mockert, Institute of Statistics, Karlsruhe Institute of Technology (KIT), Karlsruhe, Germany}
\corremail{fabian.mockert@kit.edu}
\presentadd{Institute of Statistics, Karlsruhe Institute of Technology (KIT), Blücherstr.17, 76185 Karlsruhe, Germany}
\fundinginfo{European Union: ERC, ASPIRE, 101077260, Helmholtz Association: VH-NG-1243 Young Investigator Group 'SPREADOUT', KIT Centre MathSEE, Vector Stiftung: Young investigator Group 'Artificial Intelligence for Probabilistic Weather Forecasting'.}

\runningauthor{Mockert et al.}
\begin{document}

\title{\vspace{-1.1cm}Windows of opportunity in subseasonal weather regime forecasting: A statistical-dynamical approach\vspace{-0.8cm}}

\author[1,2\authfn{1}]{Fabian Mockert}
\author[1,3]{Christian M. Grams}
\author[4,5,6]{Sebastian Lerch}
\author[1]{Julian Quinting}
\affil[1]{Institute of Meteorology and Climate Research Troposphere Research (IMKTRO), Karlsruhe Institute of Technology (KIT), Karlsruhe, Germany}
\affil[2]{now at: Institute of Statistics, Karlsruhe Institute of Technology (KIT), Karlsruhe, Germany}
\affil[3]{now at: Federal Office of Meteorology and Climatology, MeteoSwiss, Zurich-Airport, Switzerland}
\affil[4]{Institute of Statistics, Karlsruhe Institute of Technology (KIT), Karlsruhe, Germany}
\affil[5]{Heidelberg Institute for Theoretical Studies, Heidelberg, Germany}
\affil[6]{now at: Department of Mathematics and Computer Science, University of Marburg, Marburg, Germany}
\maketitle
\vspace{-0.4cm}
\begin{abstract}
\small
The Madden-Julian Oscillation (MJO) and stratospheric polar vortex (SPV) are prominent sources of subseasonal predictability in the Extratropics. With relevance for European weather it has been shown that the joint interaction of MJO and the SPV can modulate the preferred phase of the North Atlantic Oscillation (NAO) and the occurrence of weather regimes. However, improving extended-range numerical weather prediction (NWP) at three-week lead times remain under-explored. This study investigates how MJO and SPV phases affect Greenland Blocking (GL) activity and integrates atmospheric state information into a neural network to enhance week-three weather regime activity forecasts.

We define a weather regime activity metric using ECMWF reanalysis and reforecasts. In reanalyses we find increased GL activity following MJO phases 7, 8 and 1, as well as weak SPV phases, indicating climatological windows of opportunity in line with previous studies. However, ECMWF forecast skill improves only in MJO phases 8 and 1 and weak SPV phases, identifying somewhat different model windows of opportunities. Next we explore using these findings in post-processing tools. Climatological forecasts based on MJO/SPV-NAO relationships provide a purely statistical approach to extended-range GL activity forecasting, independent of NWP models. Notably, MJO conditioned climatological forecasts show clear signals when evaluated against observed GL activity.

Statistical-dynamical models, using neural networks that combine historical atmospheric state data with NWP-derived weather regime metrics improve weather regime activity forecasts across all regimes considered, achieving an absolute accuracy increase of 2.9\% in forecasting the dominant weather regime compared to ECMWF. This is particularly beneficial to Blocking in the European domain, where NWP models often underperform. Atmospheric conditioned and neural network forecasts serve as valuable decision-support tools alongside NWP models, enhancing the reliability of subseasonal-to-seasonal (S2S) predictions.
\keywords{weather regimes, windows of opportunity, Madden-Julian Oscillation, Stratospheric Polar Vortex, neural networks}
\end{abstract}

 
\section{Introduction}\label{sec:Introduction}
The demand for accurate subseasonal to seasonal forecasting -- extending beyond 10 days and up to 30 days \citep{White2017} -- has grown over the past decade, driven by socio-economic needs. Decision-makers in sectors such as public health, agriculture, and energy rely on accurate forecasts several weeks ahead, as briefly reviewed with the following examples: In the United Kingdom, winter cold spells increase hospital admissions due to respiratory diseases \citep{Charlton-Perez2019, Elliot2008}. Improved drought forecasting can save millions of dollars by mitigating crop shortages \citep{Salient2023}. Additionally, large-scale weather patterns can serve as indicators of peak winter energy demand, placing immense stress on the power systems \citep{Bloomfield2020, Millin2024}. \citet{Bloomfield2021b} show that in the context of renewable energy forecasting, pattern-based methods yield higher forecast skill than grid-point based methods for extended-range lead times > 12 days.

Large-scale quasi-stationary, recurrent, and persistent weather patterns are commonly referred to as weather regimes \citep{Michelangeli1995, Vautard1990}. They represent the complex large-scale circulations through a finite set of states, facilitating the interpretation and categorisation of the prevailing or forecasted atmospheric flow. These regimes are associated with distinct impacts on surface variables, including 2-m temperature, 10-m wind and radiation \citep{Grams2017, Beerli2019}. Often the weather regimes are identified on the basis of the 500-hPa geopotential height field (Z500). For the European region, a year-round definition of seven weather regimes has been established \citep{Grams2017} which can be formalised with a seven-dimensional normalised weather regime index. Two of the seven weather regimes, Zonal Regime and Greenland Blocking, closely resemble the positive and negative phase of the North Atlantic Oscillation (NAO), respectively. The forecast skill of Greenland Blocking and Zonal Regime is generally the best among the seven weather regimes, whereas the European and Scandinavian Blocking regimes are the worst predicted \citep{Bueler2021, Osman2023}. These three blocking regimes are especially important for the prediction of renewable power generation in winter as periods of low wind and solar power output occur mainly within these regimes. Greenland Blocking is further connected with cold temperature anomalies, leading to increased heating demand \citep{Mockert2023}. Improvements in the predictions for these three regimes, along with the other four regimes, in winter would greatly benefit the energy sector, particularly energy trading companies and transmission grid operators. 

Previous studies suggest links between the atmospheric state at initialisation time and the occurrence of weather regimes weeks later \citep[e.g.,][]{Cassou2008, Baldwin2001}. Though much weaker, these connections are reflected in the skill of numerical weather prediction models in forecasting these regimes \citep[e.g.,][]{Vitart2017b}. Two major atmospheric modes of variability influencing Greenland Blocking (or NAO--) are the Madden Julian Oscillation (MJO, \citep{Madden1971}), and the stratospheric polar vortex (SPV). 
\citet{Cassou2008} and \citet{Lin2009} showed a lagged connection between MJO phases at initialisation time and the preferred flow pattern in the North Atlantic-European region. The occurrence of NAO-- is anomalously high two weeks after MJO phases 6--8, while NAO+ is anomalously more frequent following MJO phases 3 and 4. Similarly, although considerably weaker than in reanalysis, reforecasts by the European Centre for Medium-Range Weather Forecasts show positive NAO forecasts for lead times of 11--15 days following phase 3 and negative NAO forecasts following phase 7 \citep{Vitart2017b}. \citet{Ferranti2018} describe an asymmetric connection of the MJO to forecasts of the NAO phases. Forecasts of NAO-- tend to have more (less) skill when the MJO is active (inactive) at initialisation time, whereas little sensitivity of the forecast skill to the MJO activity at initial time is found for the NAO+ phase. Thus, despite the shown MJO teleconnection to Europe, numerical weather prediction still struggles exploiting this potential source of extended-range predictability. A potential reason, is the multi-scale interaction with other modes of subseasonal variability and the chaotic growth of error \citep{Roberts2023}.

Most importantly the relation between the MJO and European weather regimes is further influenced by the Quasi Biennial Oscillation (QBO), El Niño-Southern Oscillation (ENSO), and SPV. The QBO, the leading mode of the interannual variability in the tropical stratosphere, involves a reversal of the mean-zonal wind direction, alternating between easterly (EQBO) and westerly (WQBO) phases approximately every 28 months \citep{Feng2019, Baldwin2001}. The MJO-NAO link (tendency of NAO+/NAO-- following MJO phases 3/7 as described by \citet{Vitart2017b}) is more pronounced and persistent during the WQBO, whereas it is weaker and less statistically significant during the EQBO \citep{Feng2019}. ENSO modulates the MJO-NAO signal such that the NAO+ following MJO phases 1--5 is strongly enhanced during El Niño years and suppressed during La Niña years. Conversely, NAO-- following MJO phases 7--8 is most pronounced during La Niña years and suppressed during El Niño years \citep{Lee2019}. \citet{Roberts2023} analyse the performance of ECMWF forecasts and conclude that the model underestimates the MJO-NAO connection and fails to reproduce the modulation of the MJO-NAO connection by ENSO. Finally, \citet{Kent2022} indicate that improvements in monthly NAO forecasts are hard to achieve, even with a perfect MJO forecast and MJO-NAO relationship, but they suggest that they are potentially achievable from currently unknown sources of skill. 

Greenland Blockings (NAO--) occur with an enhanced probability of up to two months following weak SPV conditions due to persistent anomalies in the troposphere \citep{Baldwin2001, Beerli2019}. Further, \citet{Domeisen2020} state that sudden stratospheric warmings (SSWs) are also frequently followed by an Atlantic Trough in the weeks after. Greenland Blocking is most likely if a blocking situation over western Europe and the North Sea prevailed, while the Atlantic Trough is most likely when Greenland Blocking was present around the SSW onset.

The connection between the atmospheric state at initialisation time and the frequency of occurrence of weather regimes, as well as the enhanced forecast skill after forecast initialisation in certain atmospheric states, indicates that the phases of the MJO and SPV could potentially be considered as windows of opportunities for subseasonal predictions. Window of opportunities can generally be defined as specific atmospheric states in which either the frequency of occurrence of an event is increased, referred to as a climatological window of opportunity, or the forecast of an event is improved, referred to as a model window of opportunity (following the definition of \citet{Specq2022} applying window of opportunities on the MJO and heavy tropical precipitation events). 

This study explores windows of opportunities linking the MJO and SPV to weather regimes during the extended winter period (November to March). A key objective is to determine whether the established links between these atmospheric modes of variability and weather regimes, typically observed up to two weeks, extend into forecast week three. To explore this, we investigate whether a climatological forecast, informed by the atmospheric state at initialisation, can provide meaningful indications of weather regime occurrence beyond week two. Additionally, we assess whether statistical-dynamical models, particularly neural networks that integrate prior knowledge and numerical weather prediction outputs, enhance weather regime forecasts.

Before answering these questions, data and methods are introduced in Section \ref{sec:DataMethod}. This study provides a detailed analysis of window of opportunities conditioned on the MJO and SPV for predicting the activity of Greenland Blocking on a lead time of three weeks (Section \ref{chap:resultWoO}). That information is then further used to provide an atmospheric-based climatology conditioned on the state of the MJO or SPV as a tool alongside a numerical weather prediction model (Section \ref{chap:resultABC}). In a final step, all information from the numerical weather prediction and the atmospheric state at initialisation time are joined by a neural network to improve the forecasts of the activity of weather regimes (Section \ref{chap:resultNeuralNets}). Section \ref{sec:Discussion} summarises and discusses the main findings.


\section{Data and methods}\label{sec:DataMethod}
\subsection{ECMWF reforecast and reanalysis}
This study utilises sub-seasonal to seasonal reforecast data of the ensemble prediction system of the European Centre for Medium-Range Weather Forecasts (ECMWF), provided through the Subseasonal-to-Seasonal (S2S) Prediction Project Database \citep{Vitart2017}. To increase the number of forecast initial dates, we include forecasts from two consecutive model cycles, Cy46R1 and Cy47R1 \citep{ECMWF2018}. These reforecasts are initialised twice a week (Mondays and Thursdays) from ERA5 reanalysis data \citep{Hersbach2020} for the past 20 years and consist of 11 ensemble members. The forecasts cover a forecast lead time of 0--46 days at a native horizontal grid spacing of 16\,km up to day 15 and 32\,km from day 15 onwards. Forecast data were remapped from their native resolution to a regular latitude-longitude grid of 1$\times$1\textdegree grid spacing.
The two model cycles were operational from June 11, 2019, to May 11, 2021, with a cycle change from Cy46R1 to Cy47R1 on June 30, 2020. As a result, for the period between May 11 and June 30 reforecasts are only available from Cy46R1. Our reforecast dataset comprises a total of 4,000 initial dates, each with 11 ensemble members. For forecast evaluation, we treat ERA5 as a "perfect ensemble member" by aligning its data with each initialisation date and lead time \citep[cf.][]{Wandel2024}. Additionally, ERA5 data is remapped from its native grid to match the reforecast grid spacing. Whenever possible (subject to predictor selection constraints), we enhance neural network training by incorporating not only the 4,000 reforecast dates but also daily ERA5 data from 1979--2021.

\subsection{Weather regimes}
In this study, we identify the seven year-round North Atlantic-European weather regimes introduced by \citet{Grams2017} in ERA5 reanalysis as described in \citet{Hauser2023, Hauser2023pre} and in the reforecasts following the approach of \citet{Bueler2021} and \citet{Osman2023}. These weather regimes (WR) represent the most common large-scale circulation patterns in the North Atlantic-European region (30--90\textdegree N, 80\textdegree W--40\textdegree E). In brief, we conduct an empirical orthogonal function (EOF) analysis of 6\,hourly (1979--2019), 10-day low pass filtered (filter-width of 20\,days, hence $\pm10$\,days), seasonally normalised geopotential height anomalies (Z500A, relative to 91-day running mean climatology) within the domain of the weather regimes. We then apply a k-means clustering algorithm on the first seven EOFs and set $k=7$. These seven clusters represent the seven distinct weather regimes (for a visualisation of the weather patterns, see Figure S1 in \citet{Mockert2024}), with three cyclonic (Atlantic Trough (AT), Zonal Regime (ZO), and Scandinavian Trough (ScTr)) and four anticyclonic regime types (Atlantic Ridge (AR), European Blocking (EuBL), Scandinavian Blocking (ScBL), and Greenland Blocking (GL)). The ZO and GL regimes correspond to the positive and negative phases of the NAO, respectively.

The projection of instantaneous anomalies onto the mean regime patterns, whether in reanalysis or reforecast, is determined by a seven-dimensional weather regime index (IWR) following the approach of \citet{Michel2011}. As this study relies on the dataset computed in \citet{Mockert2024}, we refer the interested reader to their publication for a detailed explanation and only briefly outline the computation of the IWR. Here, we specifically use the Z500 bias-corrected IWR reforecasts and reanalysis from \citet{Mockert2024}:
\begin{enumerate}
    \item Compute Z500 anomalies for reforecasts/ERA5 relative to a 91-day running mean model/ERA5 climatology (1999-2015 and 1979-2019, respectively).
    \item Apply low-pass filtering and normalisation to obtain filtered and standardised Z500 anomalies.
    \item Project filtered and standardised Z500 anomalies onto seven cluster mean Z500 anomalies of distinct weather regimes
    \item Compute weather regime index by normalising the projection against its climatological mean and standard deviation
\end{enumerate}

\subsection{Weather regime activity}
This study focuses on the predictability of weather regimes at three weeks forecast lead time. Due to the reduced skill of daily weather regime forecasts at this lead time \citep{Bueler2021} and the practical needs of decision-makers, we consider weekly aggregated forecast information based on daily data. To achieve this, we introduce the concept of weather regime activity, which quantifies the presence of a specific weather regime over a given period.
The weather regime activity is defined in two complementary ways. The \textit{weekly mean weather regime activity} ($WRact_{mean}$), suited for deterministic decision-making, considers a weather regime active in ERA5 or an ensemble member ($WRact_{mean}=1$) in the respective week, if the weekly mean of daily IWR exceeds a predefined activity threshold of IWR$_{min}=1$, otherwise $WRact_{mean}=0$. Applied to the ensembles forecast across 11 ensemble members, $WRact_{mean}$ takes values between 0 and 1 in increments of 1/11, representing a probability in the ensemble. In contrast, the \textit{aggregated daily weather regime activity} ($WRact_{agg}$) provides a more detailed view by measuring the fraction of days within a given week where the daily IWR exceeds IWR$_{min}$. Applied to ERA5 or an ensemble member, this results in values between 0 and 1, with discrete steps of 1/7, representing the fraction of days per week with active regime conditions.
When applied to ensemble forecasts across 11 ensemble members, $WRact_{agg}$ yields values between 0 and 1 in steps of 1/77, allowing for a probabilistic perspective, representing the fraction of days per week with active regime conditions in the ensemble.

\begin{figure}[!h]
    \centering
    \includegraphics[width=0.8\linewidth]{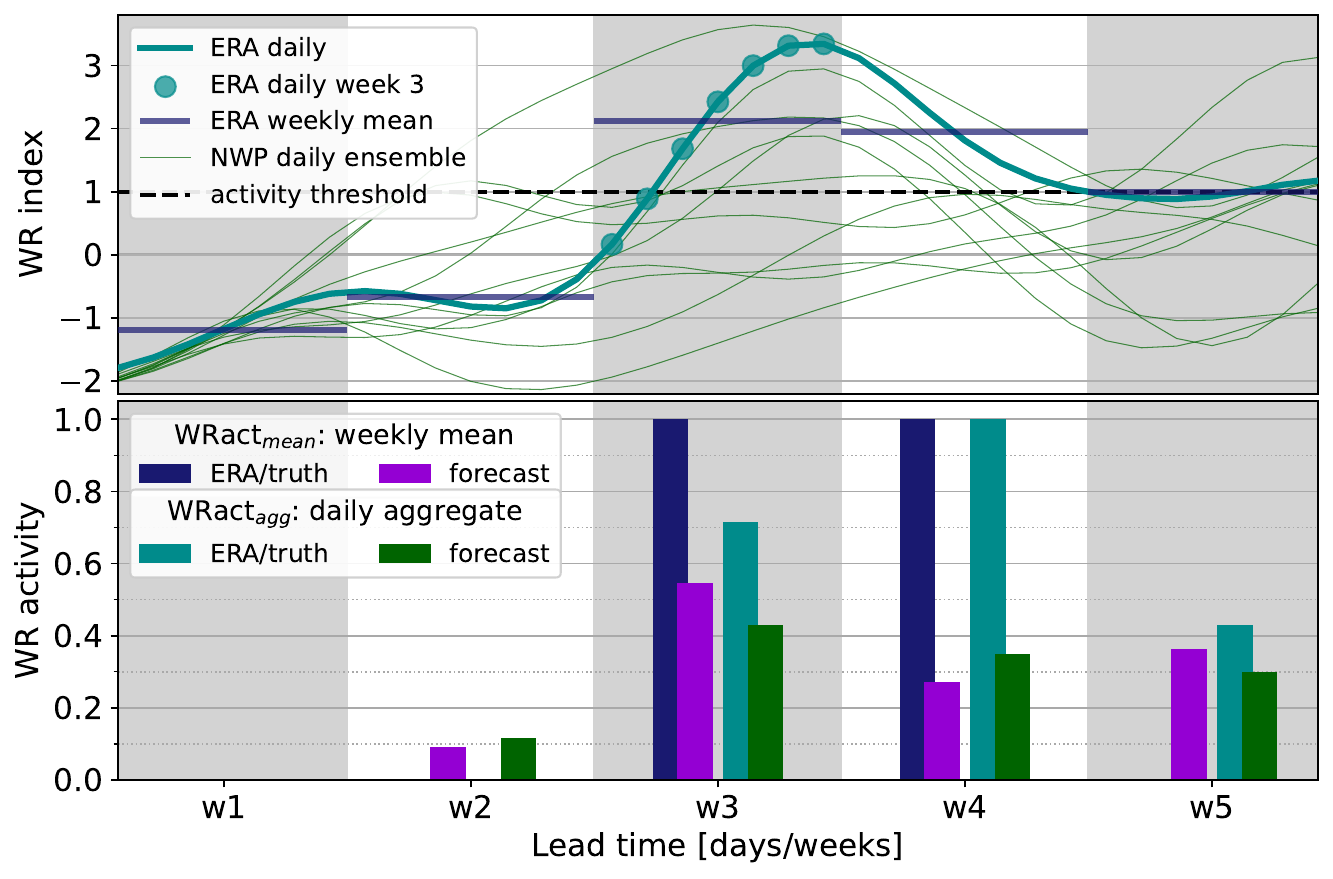}
    \caption{Visualisation of $WRact_{mean}$ and $WRact_{agg}$ by using the weather regime index. The upper panel shows the ensemble forecast (green lines) and the perfect member (ERA5, turquoise line) of the daily IWR. Further, the dark blue horizontal lines indicate the mean weekly IWR in ERA5 and the turquoise dots the daily IWR in ERA5 for forecast week 3. The latter is used for the computation of the $WRact_{agg}$ by comparing them against the activity threshold (black dashed line). In the lower panel, for each forecast week, the $WRact_{mean}$ (dark blue bars) and $WRact_{agg}$ (turquoise bars) in ERA5 and in forecasts (purple and green bars respectively) are shown. By definition $WRact_{mean}$ (dark blue bars) for ERA5 takes values of either 0.0 or 1.0.
    }
    \label{fig:WRactivitygeneration}
\end{figure}
Figure \ref{fig:WRactivitygeneration} illustrates the computation process. Seven consecutive daily IWR values from ERA5 reanalysis, shown as turquoise circles in the upper panel, are averaged to compute the weekly mean IWR (blue circles). This value (here 2.1 in week 3) is then compared to the activity threshold IWR$_{min}$ (black dashed line, here 1.0), determining whether the regime is considered active (1.0) or inactive (0.0). The corresponding $WRact_{mean}$ is represented as blue bar in the lower panel. The $WRact_{agg}$ follows a similar approach, but instead assesses each daily IWR value individually against IWR$_{min}$. The fraction of days exceeding IWR$_{min}$ defines the $WRact_{agg}$, visualised by the turquoise bar in the lower panel.

The same procedure applies to ensemble forecasts, where the IWR forecast for each ensemble member, initialised at the beginning of the shown period, is processed in the same manner (green lines in the upper panel of Figure \ref{fig:WRactivitygeneration}). The resulting $WRact_{mean}$ for the ensemble is displayed as purple bars in the lower panel, while the $WRact_{agg}$ is shown in green. This methodology enables a comprehensive assessment of weather regime activity across both, reanalysis and forecasts. Since this process is applied individually to each weather regime, multiple regimes can be considered active simultaneously, reflecting the inherent variability in large-scale atmospheric patterns.

To further simplify the interpretation of weather regime activity, we introduce the maximum weather regime activity ($WRact_{max}$), which assigns a single dominant regime to each day independent of an $IWR$ threshold. This is achieved by identifying the weather regime with the highest $WRact_{agg}$ within a given period. In case multiple weather regimes share the highest $WRact_{agg}$ at a time, the regime persisting the longest is chosen. By designating the most active regime, $WRact_{max}$ provides a practical and intuitive summary, ensuring that each atmospheric scenario is linked to the prevailing large-scale circulation pattern. With this definition, we aim to offer a simplified yet informative representation of regime dominance, making it accessible to users beyond the meteorological sector, such as decision-makers who rely on large-scale weather variability information.

\subsection{Contingency table and metrics}\label{chap:methodscontingency}
We define windows of opportunity and validate the $WRact_{max}$ on the basis of contingency tables. We utilise both, traditional and probabilistic contingency tables (following the methodology in \citet{Gold2020}), as well as verification metrics which are derived from these.
\begin{table}[h]
    \centering
    \begin{subtable}{0.48\textwidth}
        \centering
        \begin{tabular}{|c|c|c|}
        \hline
        Forecast & \multicolumn{2}{c|}{\cellcolor{white}Weather regime active/inactive} \\
        \cline{2-3}
         & Yes & No \\
        \hline
        Yes & \makecell{Hit / True \\ Positive (TP)} & \makecell{False alarm /\\ False Positive (FP)} \\   
        No  & \makecell{Miss /False \\ Negative (FN)} & \makecell{Correct rejections /\\ True Negative (TN)} \\
        \hline
        \end{tabular}
        \caption{Traditional contingency table.}
    \end{subtable}
    \hfill
    \begin{subtable}{0.48\textwidth}
        \centering
        \begin{tabular}{|c|c|c|}
        \hline
        {\cellcolor{white}Forecast} & \multicolumn{2}{c|}{\cellcolor{white}Weather regime active/inactive} \\
        \cline{2-3}
         & Yes & No \\
        \hline
        Yes & $E_{11}=\mathbf{p}\cdot\mathbf{k}$ & $E_{12}=\mathbf{p}\cdot(\mathbf{1}-\mathbf{k})$ \\   
        No  & $E_{21}=(\mathbf{1}-\mathbf{p})\cdot\mathbf{k}$ & $E_{22}=(\mathbf{1}-\mathbf{p})\cdot(\mathbf{1}-\mathbf{k})$ \\
        \hline
        \end{tabular}
        \caption{Probabilistic contingency table.}
    \end{subtable}

    \caption{General classification terms of (a) the traditional 2$\times$2 contingency table and (b) probabilistic equations for the occurrence of weather regime activities. $\mathbf{p}$ represents a vector of probabilities for a weather regime activity to occur and $\mathbf{k}$ is a vector of binary indicators whether the weather regime activities have occurred at the given dates (see Section \ref{chap:methodscontingency}). The operator "$\cdot$" denotes the inner product of two vectors and "$\mathbf{1}$" the unit vector. The tables are adopted by \citet{Gold2020} and modified.}
    \label{tab:contingencytable}
\end{table}

A traditional contingency table (Table \ref{tab:contingencytable}a) represents the relationship between forecasted and observed categorical events (e.g., whether a weather regime is active or not). The contingency table classifies outcomes into four categories: Hit (or True Positive, TP), where weather regime activity is forecasted and the weather regime activity occurs; Miss (or False Negative, FN), where no weather regime activity is forecasted but a weather regime activity occurs; False Alarm (or False Positive, FP), where weather regime activity is forecasted but no weather regime activity occurs; and Correct Rejection (or True Negative, TN), where no weather regime activity is forecasted and no weather regime activity occurs.
These categories form the basis for several verification metrics, which are detailed in Table \ref{tab:verificationmetrics}. The key metrics include Base Rate, representing the overall occurrence of a weather regime activity; Forecast Rate, which is the proportion of forecasts predicting a weather regime activity; Hit Rate (also known as Recall or Accuracy), which is the proportion of actual weather regime activities correctly predicted; False Alarm Rate, the proportion of incorrect predictions of weather regime activities; Peirce Skill Score, which is the difference between the hit rate and the false alarm rate; Precision, representing the proportion of true positive forecasts out of all positive predictions; and F1-Score, the harmonic mean of precision and Hit Rate, balancing both metrics.
\begin{table}[h]
    \centering
    \begin{tabular}{|>{\raggedright\arraybackslash}m{3.5cm}|m{3cm}|}
    \hline
    \textbf{Metric} & \textbf{Formula} \\
    \hline
    Base Rate & \( \textrm{BR} = \frac{TP + FN}{TP + FN + FP + TN} \) \\
    \hline
    Forecast Rate & \( \textrm{FR} = \frac{TP + FP}{TP + FN + FP + TN} \) \\
    \hline
    Hit Rate (Recall, Accuracy) & \( \textrm{HR} = \frac{TP}{TP + FN} \) \\
    \hline
    False Alarm Rate & \( \textrm{FAR} = \frac{FP}{FP + TN} \) \\
    \hline
    Peirce Skill Score & \( \textrm{PSS} = \textrm{HR} - \textrm{FAR}\) \\
    \hline
    Precision & \( \textrm{Prec} = \frac{TP}{TP + FP} \) \\
    \hline
    F1-Score & \( \textrm{F1} = 2 \times \frac{\textrm{Prec} \times \textrm{HR}}{\textrm{Prec} + \textrm{HR}} \) \\
    \hline
    \end{tabular}
    \caption{Collection of verification metrics originating from the contingency table in Table \ref{tab:contingencytable} used to identify windows of opportunity and to validate the skill of a categorical $WRact_{max}$ forecast.}
    \label{tab:verificationmetrics}
\end{table}

An alternative to the traditional contingency table is the probabilistic contingency table (Table \ref{tab:contingencytable}b), which incorporates probabilistic forecasts \citep{Gold2020}. In this framework, the vector $\mathbf{p} = (p_1, p_2, ... , p_n)$ represents the forecast probabilities for an event (weather regime active) occurring, where $n$ is the total number of dates in the sample. Each $p_i$ is the mean of the binary predictions from all ensemble members for date $i$, with $M$ being the number of ensemble members, in this study $M=11$. Specifically, each ensemble member $m$ provides a binary prediction $p_{i,m}$ (which is either 0 or 1), and the overall probability for date $i$ is calculated as
\begin{equation}
  p_i = \frac{1}{M} \sum_{m=1}^{M} p_{i,m}.
\end{equation}
Similarly, the vector $\mathbf{k} = (k_1, k_2, ... , k_n)$ represents the actual binary outcomes, where $k_i = 1$ if the weather regime activity occurs and $k_i = 0$ if it does not occur.
The scalar product of the forecast probabilities $p_i$ and the binary outcomes $k_i$ generates the probabilistic outcomes, which are analogous to the traditional contingency table.

In addition to the verification metrics derived from the contingency tables, we use the Mean Squared Error (MSE) to assess the quality of the forecasts (verifying $WRact_{agg}$). The MSE quantifies the average squared differences between the predicted values and the observed values across all data points:
\begin{equation}
    MSE = \frac{1}{n}\sum_{i=1}^n \left( y_i-\hat{y_i}\right)^2,
\end{equation}
where $y_i$ is the forecasted value and $\hat{y_i}$ is the observed value for the $i$-th data point. To compare the MSE of a forecast to that of a reference forecast, we compute the MSE score (MSES):
\begin{equation}
    MSES = 1-\frac{MSE_{forecast}}{MSE_{reference}}.
\end{equation}
Here, the MSES represents the normalised performance of the forecast relative to the reference, which in this case is the numerical weather prediction (NWP) forecast. A higher MSES indicates better forecast skill compared to the reference forecast.

\subsection{Windows of opportunity}\label{chap:methodsWoO}
In this study, we adopt the window of opportunity definition from \citet{Specq2022}. The atmospheric state at initialisation time can be characterised by several indicators, such as the phase of the Madden-Julian Oscillation or the strength of the stratospheric polar vortex.

We categorise the data according to the prevalent phase. Following the window of opportunity definition by \citet{Specq2022}, verification metrics (base rate (BR), hit rate (HR), Peirce skill score (PSS)) are then calculated for each subsample, based on the definitions in Table \ref{tab:contingencytable} and Table \ref{tab:verificationmetrics}. To assess whether the differences in the weather regime activity after active and inactive phases are statistically significant, we use confidence intervals. Specifically, we calculate a 90\% confidence interval for each verification metric during active phases using bootstrapping. This is done by resampling the dates of each phase 1,000 times with replacement. If the corresponding metric for the inactive phase falls outside this confidence interval, we consider the difference to be statistically significant.

Once the verification metrics (BR, HR, PSS) are calculated, we categorise windows of opportunities into two main types: climatological and model windows of opportunities, with the model windows of opportunities further divided into three subcategories. A climatological window of opportunity is based on the historical relationship between the atmospheric state at initialisation time and the likelihood of a weather regime to occur three weeks later as derived from ERA5 reanalysis. A certain atmospheric state is considered to represent a climatological window of opportunity if the frequency of weather regime occurrence -- referred to as the base rate -- is higher after an active phase than after an inactive phase. By identifying these climatological windows of opportunities, a forecaster can estimate the likelihood of a weather regime activity occurring simply by relying on past observational data corresponding to a given atmospheric condition at initialisation.

In contrast, a model window of opportunity takes into account both the base rate and the ability of the forecasting system (here the ECMWF reforecasts) to detect and benefit from the atmospheric signal. For a model window of opportunity to be useful, it might not be enough for the system to simply identify a higher base rate and hit rate. The system should also demonstrate a reliable forecast quality, which is where the Peirce skill score comes into play. The Peirce skill score assesses the difference between the hit rate and false alarm rate, allowing us to distinguish between forecasts that are genuinely improved and those that may have increased hit rates at the cost of more false alarms.
Within the model window of opportunity category, there are three subtypes: 
\begin{description}
    \item[Type 1:] Characterised by an increase in both the base rate and the hit rate, but without an improvement in the Peirce skill score due to an excessive false alarm rate. In this case, while the forecast may indicate a higher likelihood of the event, it comes with a higher number of false alarms, reducing overall forecast reliability.
    \item[Type 2:] Shows improvements in the base rate, hit rate, and the Peirce skill score. This subtype indicates a higher-quality forecast where the hit rate improves while minimising false alarms, resulting in a more accurate and reliable forecast.
    \item[Type 3:] Exhibit a positive anomaly in both the hit rate and the Peirce skill score compared to the neutral phase, even if this improvement cannot be explained by a climatological signal, due to no increase in the base rate. These phases are still considered valuable because they reflect an improvement in forecast performance over what would be expected without the active phase.
\end{description}

\subsection{Atmospheric-conditioned climatology}\label{chap:methodsABCforecast}
The existence of windows of opportunities under certain atmospheric conditions naturally motivates the computation of a climatology conditioned on the atmospheric state at initialisation time, which we refer to as an atmospheric-conditioned climatology. Unlike an unconditioned climatology, which averages over all historical periods, the atmospheric-conditioned climatology is computed conditioned on each atmospheric state (e.g., a specific MJO phase) individually. It provides the climatological mean of the $WRact_{agg}$ at a specific lag. Since atmospheric variables used for conditioning are sensitive to seasonality, all climatologies are computed using a 91-day sliding window. For each atmospheric state, we select all historical dates within the sliding window that meet the state condition at initialisation. From these dates, we compute the atmospheric-conditioned climatology of $WRact_{agg}$. These climatologies offer insight into how frequently weather regime activity occurs under different atmospheric conditions (see Section \ref{chap:resultABC}).

Building on the atmospheric-conditioned climatology, we construct an atmospheric-conditioned climatological forecast by selecting the corresponding atmospheric-conditioned climatology for each initialisation date based on the prevailing atmospheric condition. Although this simple, condition-based selection is straightforward and effective, it can lead to abrupt changes in forecasted $WRact_{agg}$ when transitioning between atmospheric states.
To address this issue, we introduce a sliding forecast, which smooths these transitions. This approach differs depending on whether the conditioning variable is two-dimensional (e.g., the MJO) or one-dimensional (e.g., the SPV index). Below, we detail the method for the MJO and highlight modifications required for one-dimensional variables.

The Madden–Julian Oscillation is the dominant mode of sub-seasonal atmospheric variability in the tropics \citep{Madden1971}. Unlike the commonly used real-time multivariate MJO index (RMM) by \citet{Wheeler2004}, we adopt the Outgoing Longwave Radiation (OLR) MJO Index (OMI) \citep{NOAA2025b}. The OMI is derived solely from OLR anomalies, without incorporating zonal winds at 850\,hPa and 200\,hPa. This results in a smoother temporal evolution of the MJO signal compared to RMM.
\begin{figure}[!h]
    \centering
    \includegraphics[width=0.6\linewidth]{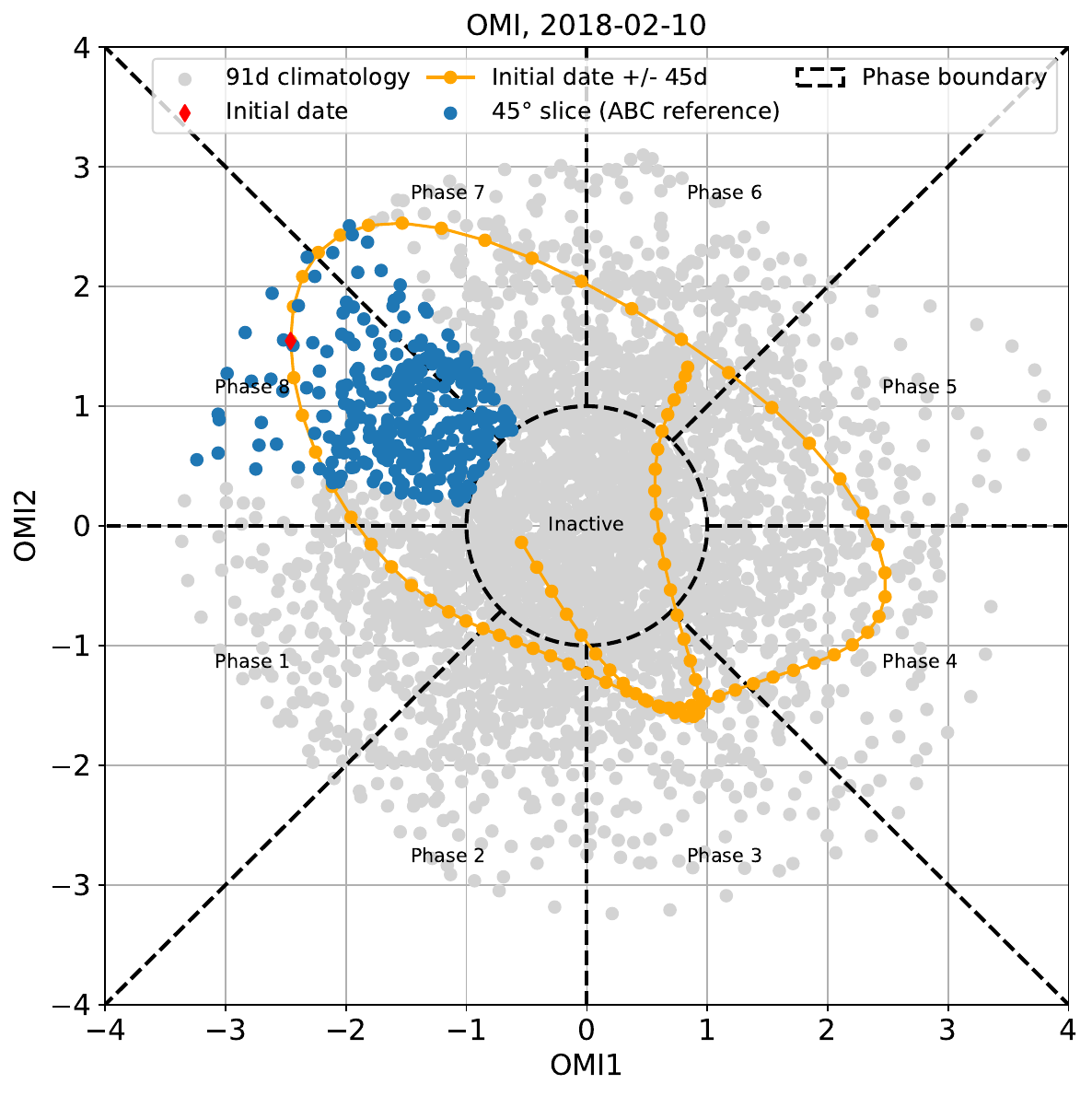}
    \caption{Phase diagram of the outgoing long-wave radiation Madden-Julian Oscillation index. The first and second principal components derived from empirical orthogonal function analysis of filtered OLR data are given on the x- and y-axis. Further the phases of the MJO are indicated with dashed lines and labelling. For visualisation how the atmospheric-conditioned climatological forecast works, one initial date, February 10, 2018, is marked with a red diamond and all values in the 45\textdegree arc used for the MJO-conditioned climatological forecast of that initialisation date (blue dots). Values in the $\pm45$ days window around the initial date (orange dots) are excluded from the data used for the computation. With grey dots, all other data points of the 91 day running climatology are indicated. For a visualisation of the OMI phase space similar to the real-time multivariate MJO index (RMM1/2), OMI1 is multiplied by -1 and OMI1 and OMI2 are exchanged with each other.
    }
    \label{fig:MJOphasediagram}
\end{figure}

The sliding atmospheric-conditioned climatological forecast for the MJO is generated as follows (illustrated in Figure \ref{fig:MJOphasediagram}): First, all historical OMI1 and OMI2 pairs (the first two principal components of the OLR anomaly empirical orthogonal function) within a 91-day sliding climatological window (grey dots) centred on the day of year of the current initialisation date (red diamond) are collected. Rather than selecting only the single MJO phase for the current date, we include all historical dates whose OMI1/OMI2 pairs fall within a 45\textdegree arc (blue dots) around the current OMI pair in the phase space. Dates with an OMI amplitude less than 1 (inner circle), representing the MJO inactive phase, are excluded. Additionally, to ensure independence from recent events all events in a windows of $\pm$ 45 days (orange dots) are removed from consideration. Finally, the mean $WRact_{agg}$ is computed from the remaining historical dates and used as the MJO-conditioned climatological forecast for the current initialisation.

For one-dimensional atmospheric variables, such as the SPV index, the sliding forecast approach is adjusted accordingly. Instead of using a 45\textdegree arc in a two-dimensional phase space, we select the 5\% of historical values closest to the current SPV value from within the 91-day sliding climatological window, excluding dates from the current seasonal cycle. The mean $WRact_{agg}$ from these selected historical dates is then used to produce the SPV-conditioned climatological forecast.

\subsection{Neural networks}\label{chap:methodsNN}
To combine information from (atmospheric-conditioned) climatological forecasts, NWP forecasts, and recent $WRact_{agg}$, we employ a statistical-dynamical approach represented with fully connected dense neural networks (NN). With these networks, we aim to identify complex relationships between predictors and enhance the predictability of $WRact_{agg}$, particularly for strong weather regime events in forecast week 3. For readers unfamiliar with machine learning techniques and terminology, we recommend the machine learning tutorials by \citet{Chase2022, Chase2023}. A fully connected dense neural network setup is selected after extensive experimentation with different architectures, depths, loss functions, and output configurations. The final models, one model per lead time and weather regime, consist of two hidden layers with 64 and 16 nodes, each using a ReLU activation function. A dropout layer with a rate of 0.2 is inserted between the hidden layers to reduce overfitting. The output layer has a single node with a sigmoid activation function, producing a probability-like output representing the $WRact_{agg}$ for a specific weather regime.

The network is trained for up to 50 epochs, but early stopping is applied to prevent overfitting. Early stopping monitors the validation loss with a patience of 10 epochs, halting training if the validation loss does not improve. Additionally, a learning rate scheduler reduces the learning rate, which starts at 0.001, by a factor of 0.5 after 5 epochs without improvement, with a minimum learning rate of $10^{-6}$. During training, the model uses a batch size of 32 and a validation split of 20\%, without shuffling the data to preserve temporal dependencies.
To robustly assess model performance, we apply 4-fold cross-validation. The 1720 data points from the extended winter period (NDJFM) are split into four consecutive subsets of 430 points each. Each fold trains on 1290 data points (data includes the validation split of 20\%) and tests on the remaining 430. To ensure results are not dependent on a single random initialisation parameter setting, ten fully independent models with identical configurations though different random initialisation parameters are trained simultaneously, and their ensemble mean is used as the final prediction. The model predicts the $WRact_{agg}$ for a single weather regime and one specific lead time, such as GL activity for forecast week 3.

To improve model performance while keeping model complexity low, we apply a stepwise feature selection procedure, restricting ourselves to the most informative predictors and keeping the signal-to-noise ratio low. The stepwise selection begins by training the model using each predictor individually and recording the mean squared error from cross-validation. In subsequent steps, the best-performing predictor from the previous step is retained, and additional predictors are tested one by one. The set of predictors with the lowest cross-validated MSE is selected. This procedure is performed separately for each lead time and weather regime, tailoring the predictor set to the specific forecast horizon and regime dynamics.

To ease interpretability, we group all available predictors into four distinct subcategories. Climatology predictors include a 91-day mean $WRact_{agg}$ climatology, day-of-year (DOY), and atmospheric-conditioned climatological forecasts based on large-scale drivers such as the MJO and SPV. Atmospheric state predictors describe the current state of the atmosphere, including indices representing large-scale circulation patterns such as the MJO and SPV, as well as sea surface temperature anomalies (SSTano) in the North Atlantic. NWP predictors capture forecast-based $WRact_{agg}$ indicators, including trends and variability across different forecast weeks. Recent weather regime predictors represent recent observed $WRact_{agg}$ and IWR from reanalysis data. A comprehensive description of these predictors is provided in Tables \ref{tab:featurelist1} and \ref{tab:featurelist2}.
\begin{table}[h]
    \centering
    \begin{tabular}{|l|c|c|c|}
        \hline
        \textbf{Predictor pools} & \textbf{$NN_{all}$} & \textbf{$NN_{NWP+WR}$} & \textbf{$NN_{noNWP}$} \\
        \hline
        Climatology & X &  & X \\
        Atmospheric State & X &  & X \\
        NWP & X & X &  \\
        Recent WR & X & X & X \\
        \hline
    \end{tabular}
    \caption{Overview of predictor pools used as input predictors for the three neural network configurations: $NN_{all}$ (all predictors), $NN_{NWP+WR}$ (NWP and weather regime predictors), and $NN_{noNWP}$ (excluding NWP predictors). An "X" indicates inclusion of the respective predictor pool.}
    \label{tab:NNpredictorpools}
\end{table}

Using these predictor subcategories, we train three separate neural networks with distinct predictor pools to explore the contributions of different predictor types (Table \ref{tab:NNpredictorpools}). The climatological neural network $NN_{noNWP}$ uses no data of the NWP model. As it does not rely on forecast data, this model is trained on a larger dataset, consisting of daily data from 1979 to 2020, though it is evaluated on the same test set as the other models for fair comparison. The NWP and weather regime neural network $NN_{NWP+WR}$ uses only NWP and weather regime based predictors, such as NWP weather regime forecast evolutions and recent past $WRact_{agg}$ and IWR from ERA. This model is trained exclusively on the reforecast period. Finally, the all-predictors neural network $NN_{all}$ combines all available predictors, integrating (atmospheric-conditioned) climatology, NWP, and recent past weather regime information. This model provides insight into the added value of integrating all sources of predictability.


\section{Results}\label{sec:Results}
We approach our investigation of windows of opportunity in subseasonal weather regime forecasting with an analysis of the modulation of occurrence and forecast quality for Greenland Blocking by different states of the MJO and SPV 
(Section \ref{chap:resultWoO}). Next, we utilise climatological knowledge of the atmospheric state to produce an atmospheric-conditioned climatological forecast for the occurrence of Greenland Blocking (Section \ref{chap:resultABC}). Finally, we integrate information from numerical weather prediction regarding $WRact_{agg}$ with knowledge about the atmospheric state at initialisation, employing a fully connected dense neural network to enhance activity forecasts and expand our analysis on other weather regimes (Section \ref{chap:resultNeuralNets}). 

The research presented here focuses exclusively on the extended winter period (November to March) and a forecast lead time of three weeks.

\subsection{Windows of opportunity}\label{chap:resultWoO}
In our setting, a window of opportunity indicates whether $WRact_{mean}$ is more likely to occur or to be forecasted three weeks after the initialisation date, conditioned on the state of the MJO or SPV phase at forecast initialisation time. To identify these opportunities, we analyse the base rate of $WRact_{mean}$ following a given phase, along with the hit rate and Peirce skill score (see Section \ref{chap:methodsWoO}).
\begin{figure}[!htbp]
    \centering
    \includegraphics[width=0.6\linewidth]{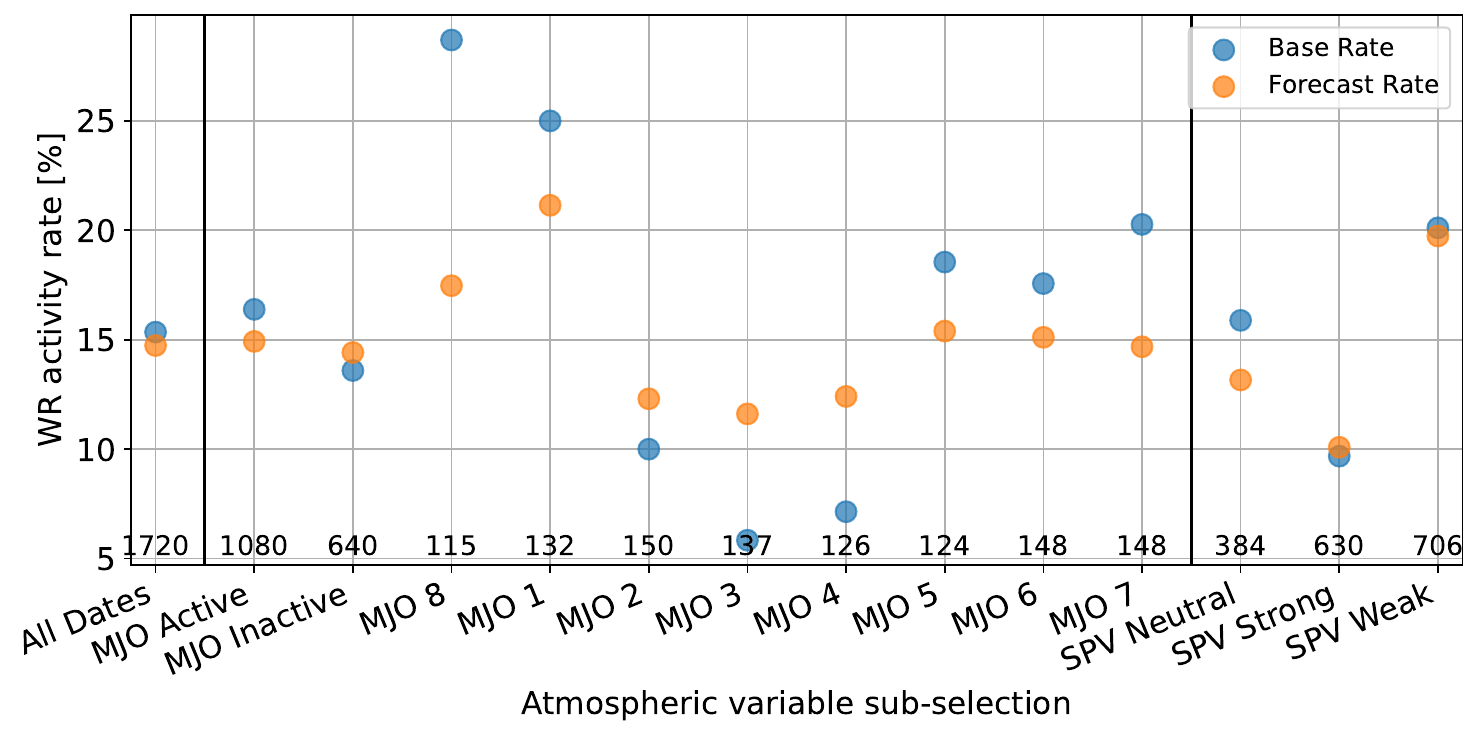}
    \caption{Base rate (blue dots) and forecast rate (orange dots) for weekly mean Greenland Blocking activity ($WRact_{mean, GL}$) in the extended winter period (November--March). These rates are split into subsets based on phases of the MJO or SPV with a lead time of three weeks. The size of the subset is indicated at the bottom of the respective phase-column.
    }
    \label{fig:baserates_GLactivity}
\end{figure}

As a preliminary step, we compare the base rate and forecast rate of weekly Greenland Blocking activity three weeks after specific states of the MJO and SPV (Figure \ref{fig:baserates_GLactivity}). For an analysis across all forecasting weeks see Figure \ref{fig:rates_GLactivity_allleads} in the Supplementary Material. On average, throughout the analysis period, the base rate and the forecast rate of $WRact_{mean, GL}$ are approximately 15\%, with the forecast rate slightly lower than the base rate, indicating a slight negative bias in $WRact_{mean, GL}$ and thus in GL frequency as found earlier \citep[e.g.,][]{Osman2023}. An analysis conditioned on the MJO phases reveals that during the inactive MJO phase, the base rate and forecast rate align closely, both at approximately 14\%. The base rate is generally higher for MJO phases 5--8 and 1, while it is lower for phases 2--4. Therefore, the phases 5--8 and 1 can be considered as climatological windows of opportunities (phases 7, 8 and 1 identify as significant for GL, according to the bootstrap test, see Section \ref{chap:methodsWoO}). Notably, forecast rates are always closer to the climatological base rate and the inactive phase than the respective base rates. This discrepancy may be attributed to the three-week forecast lead time, as extended-range forecasts tend to regress toward climatological values. 
A similar behaviour between the base rate and forecast rate can be observed for strong and weak SPV states, although the differences are negligible. The difference between the base rate and forecast rate for the neutral SPV phase is larger, which could be an indicator that the model is able to represent the occurrence of $WRact_{mean, GL}$ better when an anomalous SPV state (either strong or weak) is present at initialisation time. The positive base rate of $WRact_{mean, GL}$ after phases of weak SPV, with a significant increase compared to the neutral SPV phase, makes the weak SPV state a climatological window of opportunity. One can also consider the strong SPV state as a climatological window of opportunity but for a reduced occurrence of Greenland Blocking, which is also captured by the forecast rate.

Overall, Figure \ref{fig:baserates_GLactivity} suggests that while the forecasted frequency of weather regime activity aligns with observed occurrences, discrepancies remain in capturing the observed $WRact_{mean, GL}$ three weeks after a specific atmospheric state. Large differences exist particularly when separating by MJO phases at initialisation, with the mean frequency of $WRact_{mean, GL}$ ranging from 6\% following MJO phase 3 to 28\% following MJO phase 8. The absolute values of the base rates and forecast rates should be treated with care. The sample size for individual MJO phases, where an NWP reforecast is initialised, is small (1720 reforecasts considered across 21 winter seasons, sample size per phase is indicated at bottom of Figure \ref{fig:baserates_GLactivity}). MJO phases 7,8 and 1 as well as the weak SPV can be considered as climatological window of opportunity with significantly different base rates compared to their neutral phases, marking them as promising phases for further analysis to identify whether these climatological signals also translate into model windows of opportunities. 
\begin{figure}
    \centering
    \includegraphics[width=0.6\linewidth]{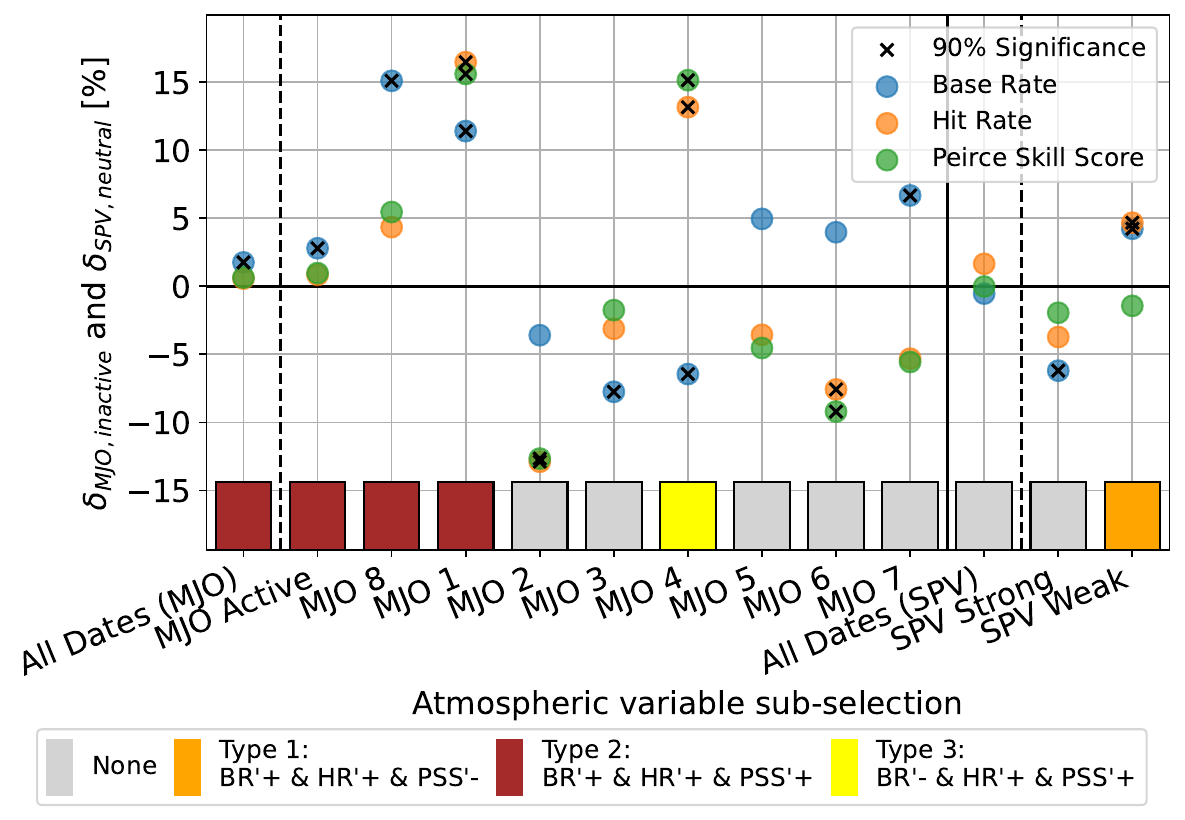}
    \caption{Indication of windows of opportunity for Greenland Blocking activity in week 3 by the base rate, hit rate and Peirce skill score anomalies and colour-coded indication. Similar to Figure \ref{fig:baserates_GLactivity} the metrics are separated by the state of the MJO and the SPV. Here, the base rate, hit rate and Peirce skill score (blue, orange, and green dots, respectively) are anomalies ($\delta$ in \%) respective to the neutral phases of the respective atmospheric variables. The combination of positive and negative anomalies (denoted with a prime) for the base rate (BR'), hit rate (HR') and Peirce skill score (PSS') are responsible for the model window of opportunity type indicated by the coloured boxes at the bottom with its legend below. Further crosses in the dots for anomalies indicate whether the rates in the specific categories are significantly different to their neutral phases with a 90\% confidence interval.
    }
    \label{fig:WoO_GLactivity}
\end{figure}

Following the approach described in Section \ref{chap:methodsWoO}, we now analyse the model windows of opportunities by computing the base rate, hit rate, and Peirce skill score for each state of the MJO and SPV. We then compute rate and skill score anomalies (denoted with a prime) relative to the inactive MJO and neutral SPV, respectively (Figure \ref{fig:WoO_GLactivity}): $R_{state}'=R_{state}-R_{ref}$, where $R\in{BR, HR, PSS}$ and $ref$ indicates either the inactive or neutral phase.

The HR' and PSS' values closely align for almost all MJO and SPV states, suggesting that the false alarm rates (FAR) in these sub-selections are similar to those of the neutral phases (not shown). The only exception is the weak SPV state during which PSS' is smaller than HR', indicating that the FAR is greater than for the neutral phase. As shown in Figure \ref{fig:baserates_GLactivity}, the BR' values are positive for MJO phases 5--8 and 1. Furthermore, for phases 8 and 1, both HR' and PSS' are positive, indicating that these phases act as a type 2 model window of opportunity, where BR', HR', and PSS' are all positive. Note that only for MJO phase 1 all three anomalies are significantly different to the neutral phase on a 90\% confidence interval.
MJO phase 4 is classified as a type 3 model window of opportunity for $WRact_{mean, GL}$ with a three-week lead time. In this case, HR' and PSS' is distinctly positive (significant difference to neutral phase), despite the BR' being negative. This suggests that, although $WRact_{mean, GL}$ is significantly less frequent following phase 4, the forecast accurately predicts these fewer occurrences, with an even lower FAR than in the neutral phase. 
The signals associated with SPV phases are less distinct, likely due to the broad classification into only three categories (strong, neutral and weak SPV) which encompasses a wide range of individual cases. Further, with the SPV index being a scalar, no information about the further temporal evolution of the index is provided. Three weeks after a weak SPV, $WRact_{mean, GL}$ is increased (positive BR'), and the forecast also predicts more frequent activity, leading to a higher HR. However, the FAR also increases, which reduces PSS'. In fact, PSS' is negative, indicating the HR/FAR ratio is worse than for the neutral SPV phase, categorising this as a type 1 model window of opportunity. In contrast, strong SPV phases are followed by less frequent $WRact_{mean, GL}$, along with a decrease in HR, PSS, and FAR, ultimately resulting in no identifiable window of opportunity type. The significant positive and negative BR' for the weak and strong SPV state, respectively, indicate that there is a strong link between the SPV and the occurrence of a weather regime activity. The negative PSS' (and therefore the higher FAR) following the weak SPV state indicate that the model has trouble predicting GL activities correctly following an extreme state of the SPV. The model shows a GL activity response too often following a weak SPV state, even when it does not realise.

In summary, the frequency of $WRact_{mean, GL}$ (base rate) varies across different MJO and SPV states. It is particularly high (significantly different to the neutral phase) for MJO phases 8 and 1, as well as after weak SPV conditions. The magnitude of the forecast rate is generally too low. MJO phases 8 and 1 emerge as type 2 model windows of opportunities, meaning that not only is the base rate increased for $WRact_{mean, GL}$ three weeks after these phases, but both the hit rate and Peirce skill score are also improved compared to the neutral phase. All three values are significantly different to the neutral phase for MJO phase 1. Forecasts initialised during a weak SPV phase show an improved hit rate, but the false alarm rate also increases compared to the neutral phase, classifying this as a type 1 model window of opportunity with significance. Finally, MJO phase 4 is a model window of opportunity for correct prediction of less GL activity. 

\subsection{Atmospheric-conditioned climatological weather regime activity forecasts}\label{chap:resultABC}
\begin{figure}[!h]
\centering
\begin{subfigure}[c]{0.8\textwidth}
\subcaption{MJO-conditioned climatology.}
\includegraphics[width=1\linewidth]{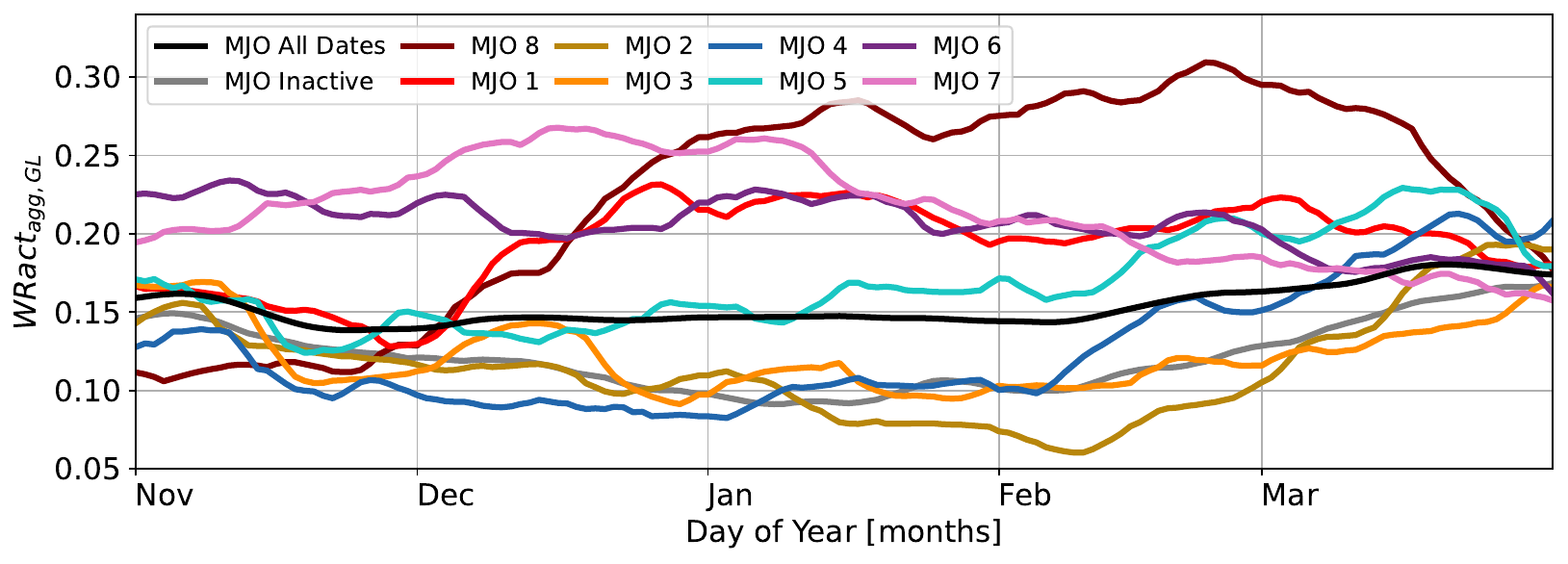}
\end{subfigure}\quad

\begin{subfigure}[c]{0.8\textwidth}
\subcaption{SPV-conditioned climatology.}
\includegraphics[width=1\linewidth]{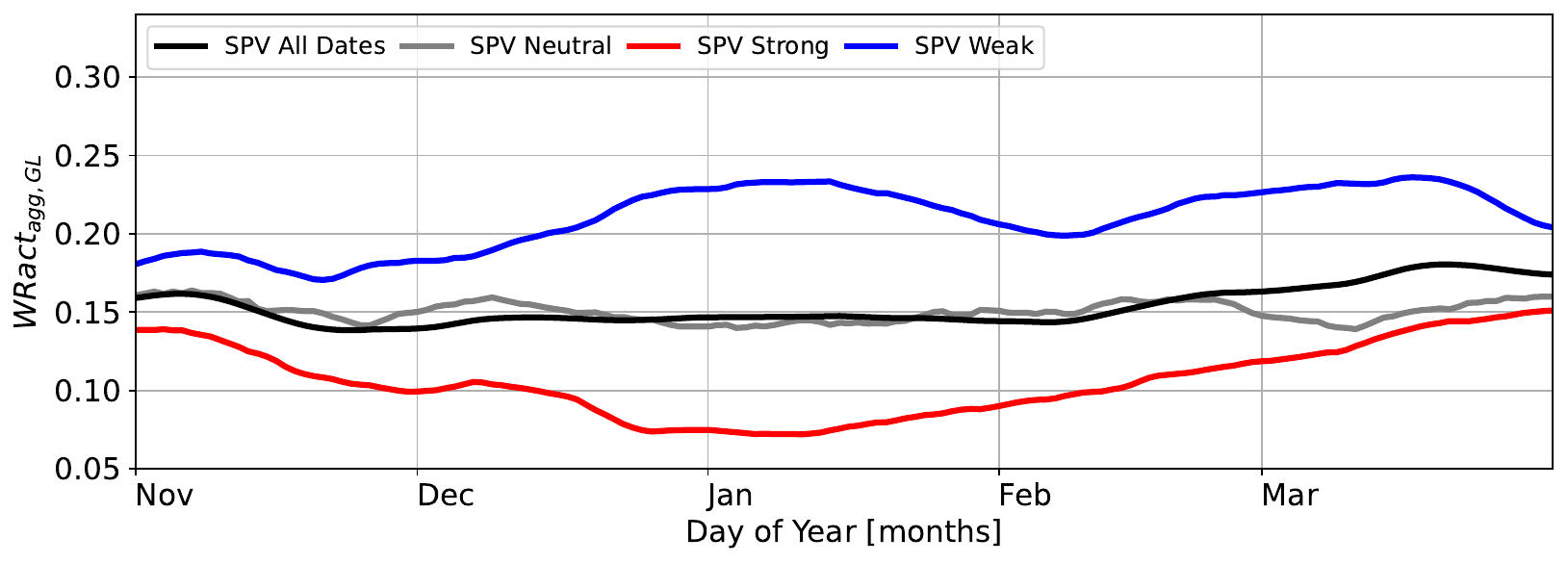}
\end{subfigure}\quad
\caption{Atmospheric-conditioned climatology for $WRact_{agg, GL}$ with a time delay of three weeks after specific states of the (a) MJO and (b) SPV for the extended winter period (November--March). The black line in both sub-figures indicate the 91-day running climatology without any atmospheric-conditioned sub-selection.
}
\label{fig:ABclimatology}
\end{figure}

The previous section and existing literature \citep{Baldwin2001, Beerli2019} suggest that the $WRact_{agg, GL}$ three weeks after forecast initialisation is influenced by the prevailing states of the MJO and SPV. To better understand this relationship, an atmospheric-conditioned climatology is introduced with a 91 day sliding window over extended winter. This climatology expresses the likelihood of $WRact_{agg}$ three weeks after a given MJO or SPV phase, demonstrating that phase-specific climatological frequencies of GL occurrence differ significantly from the overall climatology and that this has important intra-seasonal variability (Figure \ref{fig:ABclimatology}). 
During the extended winter period, the 91-day running mean of the $WRact_{agg, GL}$ fluctuates between 15 and 18\% (black lines). When categorised by MJO phases (Figure \ref{fig:ABclimatology}a), notable deviations from this baseline emerge. The most prominent signal appears for MJO phase 8, where $WRact_{agg, GL}$ increases from 11\% in November to 30\% in February (brown line). Similarly, MJO phase 1 (red line) exhibits an increase from 15\% in November to 22\% by late December, maintaining an elevated activity level of approximately 20\% until March. 
For SPV states (Figure \ref{fig:ABclimatology}b), the separation is even more pronounced. The neutral phase (grey line) largely follows the overall climatological cycle, while conditions following a weak SPV (blue line) show increased $WRact_{agg, GL}$, with two maxima reaching 24\% in January and March. Conversely, when a strong SPV (red line) is present, $WRact_{agg, GL}$ decreases, reaching a minimum of 7\% in January.

To use the atmospheric-conditioned climatology as a forecast tool, a slight modification is introduced to the computation method. Instead of strictly categorising the atmospheric-conditioned climatology by discrete phases, a sliding window approach is applied. This adjustment ensures a smoother transition between categorical phases, preventing abrupt shifts in activity forecast (see Section \ref{chap:methodsABCforecast} for further details). 

\begin{figure}[!h]
\centering
\begin{subfigure}[c]{0.5\textwidth}
\subcaption{Forecasts (0--36 days lead time) initialised three weeks (18 days) prior to GL onsets.}
\includegraphics[width=1\linewidth]{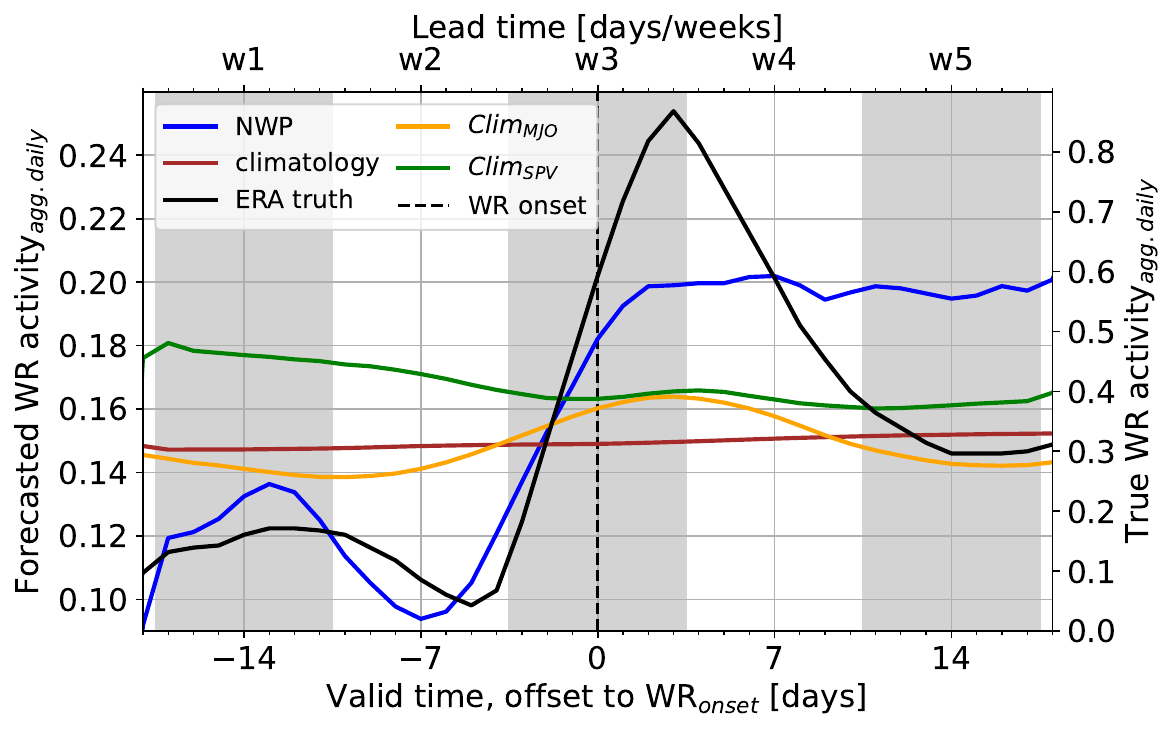}
\end{subfigure}\verb

\begin{subfigure}[c]{0.5\textwidth}
\subcaption{Forecasts at a fixed lead time of 18 days (3 weeks)  at valid times $\pm$18 days around GL onsets.}
\includegraphics[width=1\linewidth]{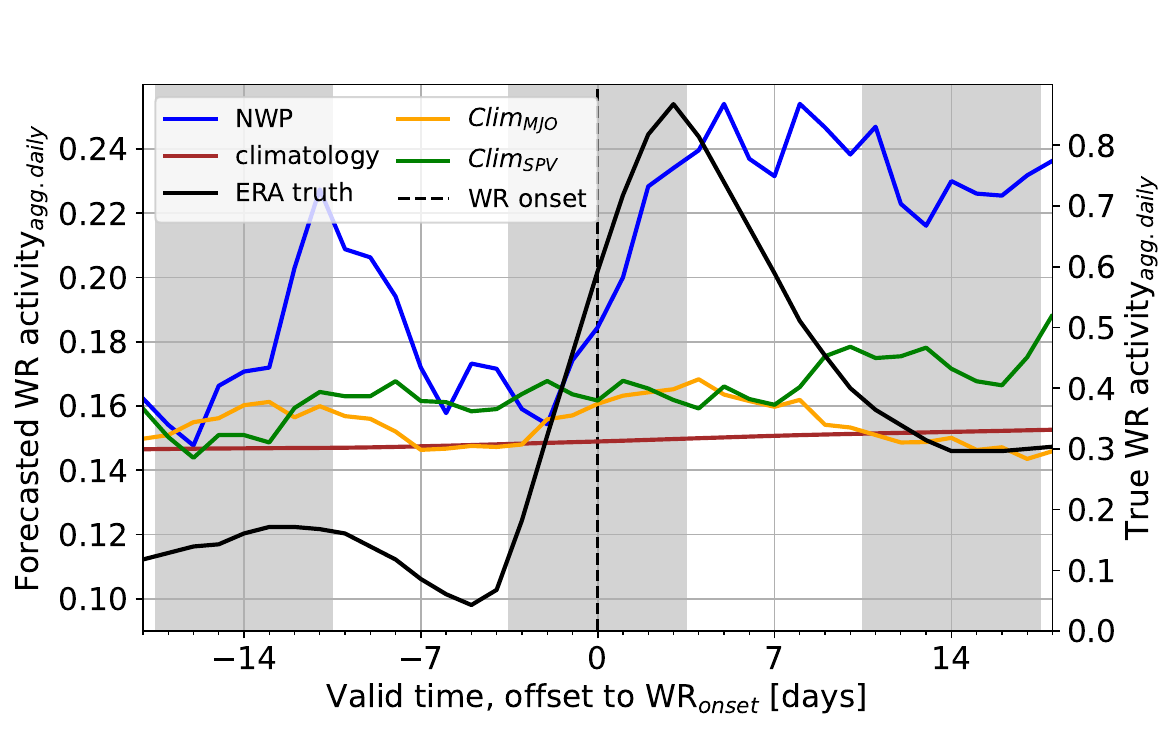}
\end{subfigure}\quad
\caption{Mean composite of different forecasts at valid times centred around observed Greenland Blocking onsets(black dashed vertical line). Due to values on different scales, forecast values for the $WRact_{agg}$ are indicated on the left y-axis, the mean of the ERA5 $WRact_{agg}$ are indicated on the right y-axis. In (a), the mean composite is generated by forecasts (0--36 day lead time) initialised three weeks (18 days) prior to GL onset, whereas in (b), the forecasts have a fixed lead time of three weeks (18 days) for each valid time $\pm$18 days around the GL onset (x-axis). For a fair comparison, between the atmospheric-conditioned climatological and NWP forecasts, we restrict the data for the computation of the mean composite to the extended winter period from 1999--2020, the period in which NWP forecasts are available. A Greenland Blocking onset is defined as the time step where the $WRact_{mean}$ for Greenland Blocking changes from 0 to 1.}
\label{fig:AroundWRonset}
\end{figure}

As a proof of concept, we select periods of active GL and analyse atmospheric-conditioned climatological and NWP forecasts leading up to these GL activities. While this approach may seem selective, as it focuses only on active GL periods, it enables a targeted examination of two key forecasting questions:
\begin{enumerate}
\item Do single forecast runs of the NWP model and atmospheric-conditioned climatological forecasts indicate the onset of $WRact_{agg, GL}$ in week 3 and the subsequent evolution of $WRact_{agg, GL}$? 
\item Do consecutively initialised week 3 forecasts show an increasing signal leading up to the $WRact_{agg, GL}$?
\end{enumerate}
To address these questions, two mean composite plots are computed (Figure \ref{fig:AroundWRonset}). 
For both analyses, the actual GL activity onset is defined as the first time when the ERA5 $WRact_{mean, GL}$ reaches one.  
For the first question, individual forecasts initialised three weeks prior to GL activity onset (or as close as possible, given the availability of NWP reforecasts) are analysed with lead times ranging from 0--36 days (Figure \ref{fig:AroundWRonset}a). For the second question, only those forecasts at a fixed lead time of three weeks (18 days) are considered. Thus, instead of a single forecast per event, consecutive forecasts leading up to the GL activity are analysed to track the signal of GL activity over time (Figure \ref{fig:AroundWRonset}b). 
The mean composite of forecasts in Figure \ref{fig:AroundWRonset} is computed centring all events around the GL onset. For the individual events see Figures \ref{fig:AroundWRonset_individual_singlefcst} and \ref{fig:AroundWRonset_individual_multifcst} in the Supplementary Material.

The true (ERA5) $WRact_{agg, GL}$ (black line, right y-axis) is identical in both sub-figures of Figure \ref{fig:AroundWRonset}. As expected, there is a clear increase in $WRact_{agg, GL}$ around the onset (vertical black dashed line), peaking three days after the onset. Additionally, a secondary local maximum appears around 12 days prior to the GL onset, indicating that in some of the scenarios, two GL events occur in close succession (e.g., GL event indices 129, 137, 181 in Figure \ref{fig:AroundWRonset_individual_singlefcst}). This secondary maximum may also indicate that the state of WR activity prior to, at, and after initialisation carries valuable predictive information. Due to the centring of the data around the activity onset, the mean composite shows actual $WRact_{agg, GL}$ reaching up to 85\%, whereas forecasts only reach up to 25\% of $WRact_{agg, GL}$ in the mean composite. Therefore, the $WRact_{agg, GL}$ forecast values are displayed on a separate y-axis for better visualisation (left y-axis).

The unconditional climatology (brown line) serves as a reference, showing no distinct signal apart from the seasonal base rate of approximately 15\% $WRact_{agg, GL}$. Comparing the MJO- and SPV-conditioned climatological forecasts (orange and green lines, respectively) to the NWP model (blue line), the NWP forecast exhibits the strongest $WRact_{agg, GL}$ signal. This result is not surprising, as the NWP model incorporates full dynamical atmospheric information, while the atmospheric-conditioned climatological forecasts rely solely on the MJO or SPV phase at initialisation. Thus, rather than comparing absolute values, the temporal changes of the forecast signals relative to the unconditional climatology are the primary focus.

For forecasts initialised three weeks prior to GL onsets (Figure \ref{fig:AroundWRonset}a), the NWP forecast shows a strong increase in $WRact_{agg, GL}$ beginning approximately seven days before onset, mirroring the observed activity trend. An additional increase appears at earlier lead times, specifically 11--17 days before onset, aligning with the secondary maximum observed in the ERA5 $WRact_{agg, GL}$. At longer lead times (7--18 days after onset), the NWP forecast suggests a prolonged high $WRact_{agg, GL}$, even when the observed activity declines. This behaviour is likely an artifact caused by forecast timing errors, where some forecasts predict $WRact_{agg, GL}$ too late (e.g., GL indices 142, 166, 202 in Figure \ref{fig:AroundWRonset_individual_singlefcst}). As a result, in the mean composite, $WRact_{agg, GL}$ appears misleadingly prolonged in the forecast.

The MJO-conditioned climatological forecast impressively captures the key characteristics of $WRact_{agg, GL}$ evolution. At earlier lead times (up to five days before onset), $WRact_{agg, GL}$ is lower than the unconditional climatology. However, from 4 days before to 10 days after onset, forecasted $WRact_{agg, GL}$ exceeds the unconditioned climatology, indicating an increased likelihood of $WRact_{agg, GL}$. This result reinforces previous findings that MJO phases modulate $WRact_{agg, GL}$ two to four weeks later, validating the teleconnection mechanism.

In contrast, the SPV-conditioned climatological forecast produces a less distinct signal. Forecasted $WRact_{agg, GL}$ remains higher than the unconditional climatology across all lead times, yet the highest values appear at early lead times, followed by a steady decline, with only a minor incline 0--7 days after the weather regime onset. One possible explanation is that the one-dimensional SPV index lacks the complexity of the two-dimensional MJO index (amplitude and phase angle). Previous studies \citep[e.g.,][]{Domeisen2020} suggest that sudden stratospheric warming events (low SPV index values) can trigger concurrent GL activity, but not all GL activities are linked to SSWs. Since the SPV-conditioned climatological forecast does not account for changes in the index over time, it merges scenarios where the atmosphere is both approaching and departing from an SSW event, leading to a diffuse climatological signal. Incorporating temporal changes in the SPV index could improve the clarity of the forecast signal and better represent the dependence between the SPV index at initialisation and the $WRact_{agg, GL}$ with a lag in time. To account for this, we introduced a 2D index that incorporates the SPV change over the past seven days. However, integrating this into the atmospheric-conditioned climatological forecast did not yield noticeable improvements (therefore not shown).

When analysing consecutive forecasts with a fixed 3-week lead time (Figure \ref{fig:AroundWRonset}b), similar signals emerge, though with distinct differences. The NWP forecast still predicts GL activity onsets, indicated with an activity increase starting for valid times two days prior to the onset. In contrast to the single forecast runs, the $WRact_{agg, GL}$ forecasts prior to the weather regime onset are already higher than climatology. The delayed signal increase compared to ERA5 (two days rather than five days prior to the onset) and the later maximum of the $WRact_{agg, GL}$ (five days rather than three days after the onset) suggest a systematic delay in predicted onset timing. Additionally, the previously active GL activity is still visible in the forecasts by the NWP (7--14 days prior to the GL onset). This could also indicate that the NWP model favours the persistence of a GL activity rather than the transition into another weather regime.  
The atmospheric-conditioned climatological forecasts continue to provide meaningful signals. The MJO-conditioned climatological forecast exhibits a clear increase relative to the unconditional climatology, beginning three days before the onset and persisting for up to ten days after the onset. The SPV-conditioned climatological forecast, while more diffuse, shows a gradual increase across the forecasted periods with all values being above climatological forecast values, supporting the notion that the SPV phase influences GL activity, albeit less distinctly than the MJO phase. 

These results demonstrate that atmospheric-conditioned climatologies offer a valuable framework for forecasting $WRact_{agg, GL}$. The MJO-conditioned climatological forecast exhibits a robust increase in $WRact_{agg, GL}$, aligning well with observed $WRact_{agg, GL}$ and confirming the role of the MJO in modulating $WRact_{agg, GL}$ two to four weeks later. The SPV-conditioned climatological forecast also indicates a weak modulation of GL activity, though it is more diffuse and less clear to interpret. While the NWP model predicts $WRact_{agg, GL}$ well, it exhibits a systematic delay of predicting the onset. Overall, these findings support the use of atmospheric-conditioned climatological forecasts as a complementary tool to dynamical NWP models. However, further refinements, particularly incorporating temporal changes in the SPV index, could enhance predictive accuracy. It is important to emphasise that these findings provide only a partial interpretation of the forecast signals, as the analysis is limited to mean composites centred around actual ERA5 $WRact_{agg, GL}$ onsets. The forecasts within this subset exhibit inherent variability, and the composites do not account for scenarios where, for example, the NWP model predicts $WRact_{agg, GL}$ that does not verify.

\subsection{Statistical-dynamical approach}\label{chap:resultNeuralNets}
Given the promising results observed in $WRact_{agg, GL}$ forecasts conditioned on the MJO and SPV, as well as the performance of the NWP model, we explore whether these forecasts --along with additional atmospheric variables-- can be combined to enhance the NWP model’s skill in predicting $WRact_{agg}$ three weeks in advance. To achieve this, we employ statistical-dynamical models using fully connected neural networks (as described in Section \ref{chap:methodsNN}) to predict $WRact_{agg}$ at week 3.

To systematically assess the impact of different information sources, we construct three neural networks based on distinct predictor pools (see Table \ref{tab:featurelist1} and \ref{tab:featurelist2} for a complete list of predictors). The "NWP and weather regime" neural network ($NN_{NWP+WR}$), combines information from the NWP model with recent $WRact_{agg}$. The "no NWP" neural network ($NN_{noNWP}$), incorporates all available predictors except those derived from the NWP model. The all-predictors neural network ($NN_{all}$) utilises the full set of available predictors, including climatological forecasts, atmospheric indicators at initialisation, NWP-based weather regime forecasts and temporal changes, and $WRact_{agg}$ history.

Each neural network undergoes a stepwise feature selection process (see Section \ref{chap:methodsNN}) to identify the most relevant predictors. Once the predictors yielding the highest skill are determined, models are retrained using k-fold cross-validation, ensuring a robust evaluation across multiple forecast instances. While this approach introduces minor information leakage -- since the predictor selection is performed on the mean over all folds before final model training on the same folds -- the effect is negligible and outweighed by the benefit of a larger testing dataset.

Our analysis begins with the predictor selection for the $NN_{all}$ in predicting $WRact_{agg, GL}$ for Greenland Blocking, followed by a comparison of the predictor order across all weather regimes.
The most influential predictor for $WRact_{agg, GL}$ in $NN_{all}$ is the NWP model’s week-three mean IWR forecast, followed closely by its $WRact_{agg}$ forecast of the NWP model for the same week, which is inherently derived from the IWR forecast (Figure \ref{fig:featureselection_GL_ABCNWP}). The second most important predictor is the NAO index at initialisation. As GL strongly correlates with the NAO--, this likely highlights the persistence of GL activity.
The next two predictors include the GL climatological occurrence and the day of the year, both are indicators that there is a seasonal cycle of the GL occurrence within the extended winter period on which the neural network can rely on. Other key predictors include the atmospheric-conditioned climatological forecasts and actual indices of the SPV and QBO, two major atmospheric modes of variability influencing the GL activity (as shown in the Introduction (Section \ref{sec:Introduction}) and in the previous results for SPV (Section \ref{chap:resultWoO} and \ref{chap:resultABC}). Interestingly, MJO-related predictors are absent in $NN_{all}$ for $WRact_{agg, GL}$, despite strong signals in prior analyses.

\begin{figure}[!h]
    \centering
    \includegraphics[width=0.8\linewidth]{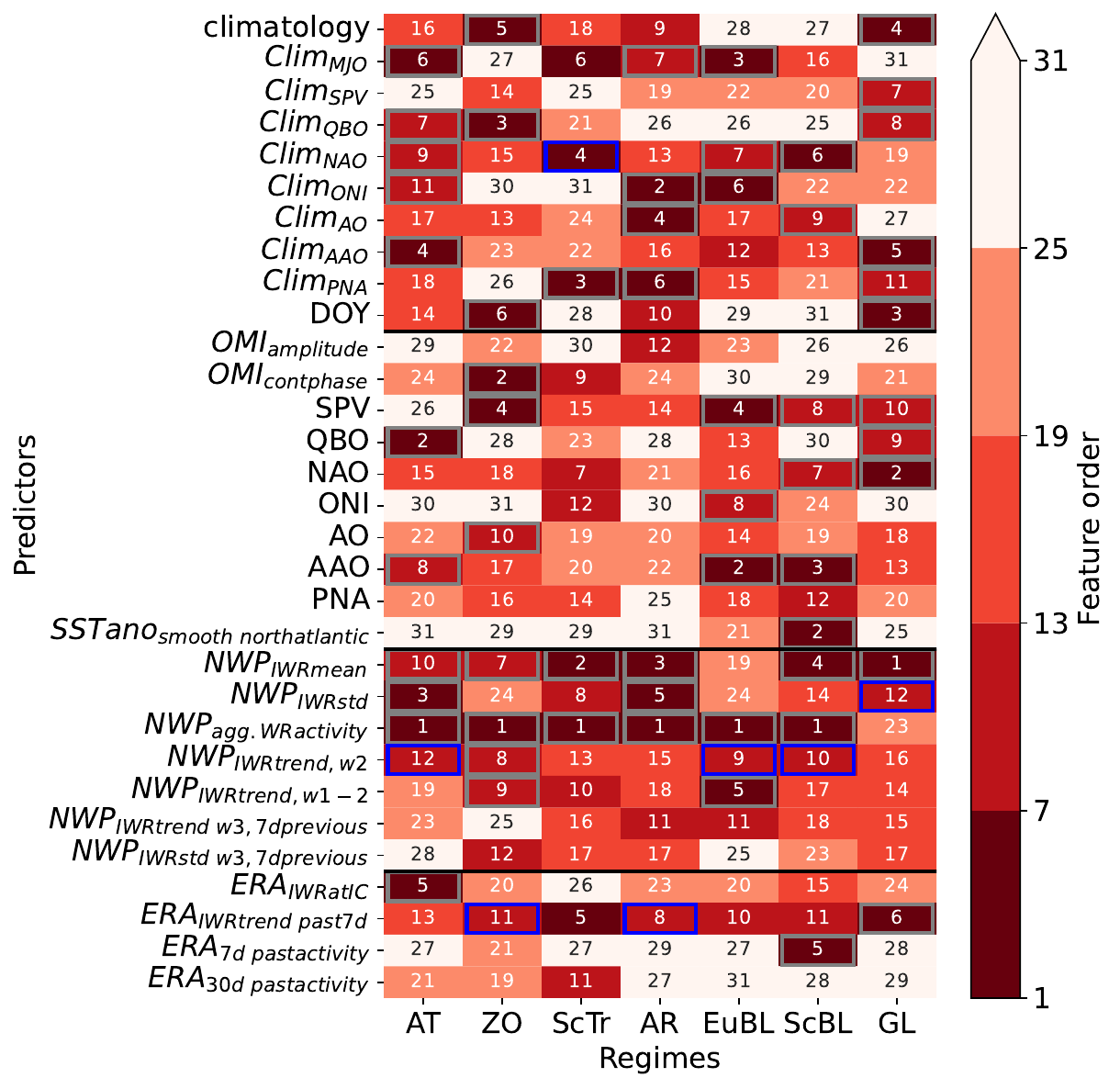}
    \caption{Summary of the predictor order for all-predictor neural networks ($NN_{all}$). The predictor order is shown for all seven weather regimes. Blue boxes indicate the last predictor which is considered due to the best MSE for the predictor combination of all predictors with a lower predictor order number (indicated with grey boxes). A detailed explanation of the predictors can be found in Table \ref{tab:featurelist1} and \ref{tab:featurelist2}.
    }
    \label{fig:featureorder_ABCNWP}
\end{figure}
Extending the predictor selection analysis to all seven weather regimes (Figure \ref{fig:featureorder_ABCNWP}) reveals that, except for GL, the NWP forecast of the $WRact_{agg}$ (target variable) remains the dominant predictor. NWP-derived predictors designed to mimic a human forecaster’s analysis -- such as the temporal change of the IWR across consecutive forecasts or its evolution in week two -- play a minor role in the neural networks. The week-two trend is the last predictor selected for AT, EuBL, and ScBL, while temporal changes from previous forecasts with the same valid time are excluded entirely.
Similar to GL, the ZO network (representing the NAO+ phase, the counterpart to GL/NAO--) includes climatology and DOY. However, here, the MJO phase, the QBO-conditioned climatological forecast, and the SPV index take precedence over climatology and DOY. An unexpected but recurrent predictor is the Antarctic Oscillation (AAO). It ranks second for EuBL, third for ScBL, eighth for AT, fourth and fifth (atmospheric-conditioned climatological forecast) for AT and GL, highlighting its potential link to $WRact_{agg}$. The AAO has been associated with its Northern Hemisphere counterpart, the Arctic Oscillation (AO), which is closely linked to the NAO. \citet{Tachibana2018} demonstrate that the AAO and AO exhibit correlations, particularly in October and February, while \citet{Song2009} find that negative AAO phases correspond to anomalously high 300 hPa geopotential heights over the North Atlantic-European region, with a lag of 25--40 days. This is consistent with the AAO being a good predictor for EuBL and ScBL in our neural networks.

The highly variable predictor selection across neural networks for different weather regimes highlights the diverse sources of predictability. Key predictors stem from the Arctic (AO-conditioned climatological forecast for AR and ScBL), the stratosphere (SPV-related predictors for ZO, EuBL, ScBL, and GL; QBO-related predictors for AT, ZO, and GL), and even the Southern Hemisphere (AAO-related predictors for AT, EuBL, ScBL, and GL; El Niño-related predictors for AT, AR, EuBL). This diversity underscores the complexity of weather regime forecasting in the North Atlantic-European region, where no single factor -- aside from NWP predictors -- dominates $WRact_{agg}$ across all regimes.

\begin{figure}[!h]
    \centering
    \includegraphics[width=0.6\linewidth]{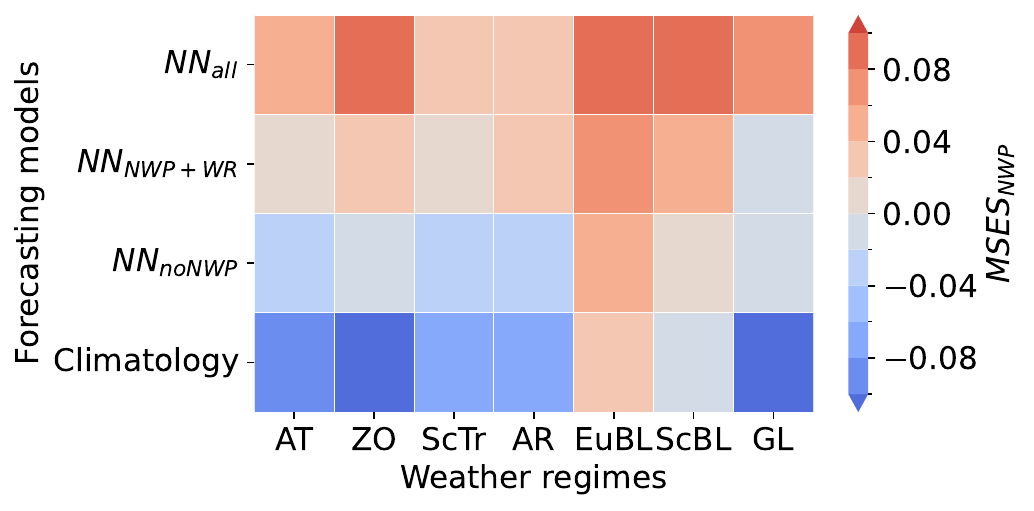}
    \caption{Performance of the different neural network and climatological forecasts in comparison to the NWP forecast for $WRact_{agg}$. As performance measure, the MSE is calculated for each model and weather regime and put into perspective against the MSE of the NWP model (MSE skill score). 
    }
    \label{fig:MSES_againstNWP}
\end{figure}
Following the predictor selection, we evaluate the neural networks' performance in  forecasting the $WRact_{agg}$ (Figure \ref{fig:MSES_againstNWP}). Comparing the MSE skill score (MSES$_\textrm{NWP}$) across all forecasts in the extended winter season provides key insights into the statistical-dynamical model performance relative to both the NWP model and climatology. The climatological forecast, based on a standard 91-day running mean of $WRact_{agg}$, serves as a baseline for assessing NWP skill. Consistent with previous research \citep[e.g.,][]{Osman2023, Bueler2021}, the NWP model struggles to outperform climatology for ScBL and, particularly, EuBL at week 3, while performing well for ZO and GL. 
All three neural network setups generally outperform the NWP model for EuBL and ScBL, with $NN_{all}$ showing the most pronounced improvements of 9.7\%. Notably, $NN_{noNWP}$ surpasses the NWP model for EuBL and ScBL, demonstrating that even without NWP-derived inputs, it achieves lower mean squared errors in $WRact_{agg}$ forecasts. Meanwhile, $NN_{NWP+WR}$, provides only minor improvements compared to the NWP model, suggesting that it mainly corrects biases rather than introducing new sources of skill.
The most striking result comes from $NN_{all}$, which consistently outperforms the NWP model across all weather regimes, achieving MSE reductions of 3.0--9.7\%, with the largest gains observed for ScBL activity. Furthermore, $NN_{all}$ is the only neural network that enhances the already high forecast skill of $WRact_{agg, GL}$. These improvements go beyond simple error reduction; as illustrated in the $WRact_{agg}$ forecasts (Figure \ref{fig:forecast_timeline}), some neural networks better capture key $WRact_{agg}$ patterns than the NWP model alone. Notably, $NN_{noNWP}$ successfully captures the prolonged GL activity during the winters of 2009/2010 and 2010/2011. Additionally, for forecasts with the largest x\% differences in predicted activity between $NN_{all}$ and NWP, $NN_{all}$ consistently achieves lower MSE, regardless of the magnitude of the difference (not shown here).
\begin{figure}[!h]
    \centering
    \includegraphics[width=0.7\linewidth]{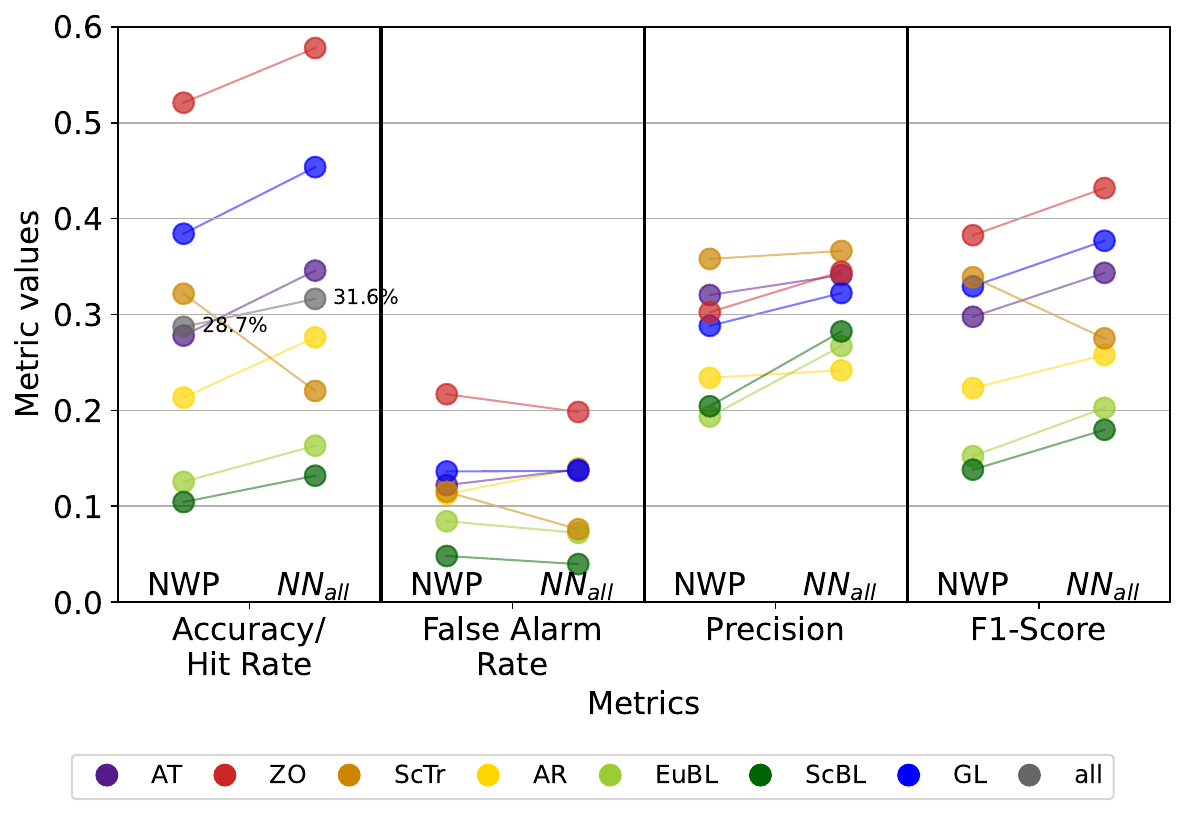}
    \caption{Verification metrics comparing the skill of the NWP model and the all-predictors neural network in predicting the $WRact_{max}$ at a lead time of three weeks. Four metrics, the hit rate, the false alarm rate, the precision, and the F1-score are visualised (x-axis). Further together with the hit rate, the accuracy for the full time series is given and indicated by percentages in text. For each metric, on the left the NWP model and on the right the all-predictors neural network ($NN_{all}$) is represented, joint by a thin line which serves as a guiding line for the reader to directly see whether the respective metric increases or decreases for the neural network in comparison to the NWP model.
    }
    \label{fig:maxWRscores}
\end{figure}

In the final stage of our analysis, we evaluate whether the improvements observed for individual $WRact_{agg}$ forecasts translate into better predictions of the $WRact_{max}$ three weeks after forecast initialisation. We determine the weather regime with the maximum predicted activity for both the $NN_{all}$ and the NWP model, then compare their performance using verification metrics derived from the contingency table (Table \ref{tab:contingencytable}a and Table \ref{tab:verificationmetrics}), including accuracy (hit rate), false alarm rate, precision, and F1-score (Figure \ref{fig:maxWRscores}). Across all metrics, a rather uniform image emerges, the $NN_{all}$ performs equal or better for all weather regimes except the ScTr. The hit rate improves for all weather regimes except ScTr, with the overall accuracy increasing from 28.7\% for the NWP model to 31.6\% for the $NN_{all}$, representing a relative improvement of 10\%. False alarm rates decrease for ZO, ScTr, EuBL and ScBL, stay the same for GL, and increase for AT and AR. The precision and F1-score improve for all weather regimes except the F1-score for ScTr, which is due to the decrease in the hit rate.

In conclusion, neural networks can extract additional predictive information beyond what is provided by the NWP model alone. While the NWP-based $WRact_{agg}$ forecast (or IWR forecast for GL) remains the most influential predictor, added value comes from atmospheric indicators such as the SPV, QBO, MJO, and AAO indices. The $NN_{all}$ not only outperforms the NWP model for individual weather regime forecasts (based on the MSE) but also achieves a 10\% relative improvement in predicting the dominant weather regime at week 3. Furthermore, the $NN_{noNWP}$, which excludes NWP-derived inputs, performs equally well or better for blocked regimes (AR, EuBL, ScBL, GL) highlighting the importance of non-NWP sources of predictability. These findings underscore the potential of hybrid forecasting approaches that integrate dynamical and statistical models to enhance sub-seasonal $WRact_{agg}$ predictions.


\section{Conclusions and discussions}\label{sec:Discussion}
Various studies documented the connection between subseasonal sources of predictability (e.g., MJO or SPV) and the large-scale circulation in the North Atlantic-European region \citep[e.g.,][]{Cassou2008, Lee2019, Lee2020, Domeisen2020, Roberts2023}. However, only few studies so far could demonstrate its practical applicability to improve extended-range NWP forecasts \citep[e.g.,][]{Scaife2022}.

In this study, we investigate the influence of the state of the MJO and SPV at initialisation on the occurrence and forecast of seven year-round North-Atlantic European weather regimes, with a focus on the activity of Greenland Blocking, three weeks later. Furthermore, we leverage information describing the atmospheric state prior to and at initialisation to enhance NWP week 3 weather regime activity forecasts using neural networks.

We compute a weather regime activity metric ($WRact_{mean}$ and $WRact_{agg}$), representing the fraction of daily weather regime index values in a given week exceeding a given threshold, based on reanalysis and reforecasts. Our findings show that Greenland Blocking $WRact_{mean, GL}$ is significantly enhanced following MJO phases 7, 8 and 1, as well as weak SPV state, suggesting a climatological window of opportunity (base rate anomaly to neutral state is positive). However, ECMWF reforecast hit rates show improvement over the respective neutral phase only in MJO phases 8 (though not significant) and 1 and in the weak SPV phase. A positive Peirce skill score in MJO phases 8 and 1 classifies them as a model window of opportunity type 2 (according to \citet{Specq2022}, with a positive anomaly in base rate, hit rate and Peirce skill score with respect to neutral state), while the weak SPV state corresponds to a model window of opportunity type 1 (base rate and hit rate anomalies to neutral state are positive and Peirce skill score anomaly is negative). Greenland Blocking is rare following the MJO phase 4, though the NWP model is performing well in correctly predicting these rare Greenland Blocking, classifying MJO phase 4 as a window of opportunity type 3 (hit rate and Peirce skill score anomalies are positive and the base rate anomaly is negative). 
We demonstrate the potential skill for predicting Greenland blocking activity in week 3 purely derived from reanalysis data by constructing an atmospheric-conditioned climatological forecast that utilises historical information of the MJO state (amplitude and phase) or SPV state to estimate the mean historical $WRact_{agg, GL}$ for the prevailing situation.
Incorporating this information -- alongside MJO, SPV, and other atmospheric variability indices -- with NWP-derived $WRact_{agg}$ metrics in a statistical-dynamical model using neural networks improves $WRact_{agg}$ forecasts across all seven weather regimes. When aggregating individual $WRact_{agg}$ forecasts to determine the dominant regime ($WRact_{max}$), the best-performing neural network achieves accuracy increases for all regimes individually, except the Scandinavian Trough. The overall accuracy increase of the neural network is 2.9\%, representing a relative improvement of 10\% compared to the NWP model.

Our findings on windows of opportunities align with previous studies on the MJO-NAO \citep{Cassou2008, Vitart2017b, Ferranti2018} and SPV-NAO \citep{Büeler2020, Baldwin2001, Beerli2019} connection. In particular, we find an increased climatological occurrence of Greenland Blocking following weak SPV states, a relationship that is well captured by the ECMWF forecast model. This supports the results of \citet{Spaeth2024}, who show that ensemble forecasts exhibit greater confidence in predicting the dominant regime -- typically Greenland Blocking -- after a weak SPV state compared to neutral or strong states. However, our results also reveal a downside: the NWP model tends to over-predict Greenland Blocking in these situations, leading to - as we can show -  frequent false alarms. Vice-versa the NWP model shows poor performance in predicting the relatively rare Greenland Blocking occurrence during strong SPV states. These limitations are likely related to the findings of \citet{Domeisen2020}, who show that the tropospheric response to a sudden stratospheric warming (i.e., a weak SPV state) can bifurcate depending on the tropospheric conditions at the time of the warming. It is plausible that the model lacks the ability to distinguish between these different pathways of stratospheric forcing.

To our knowledge, this study is the first to demonstrate how the MJO/SPV-NAO relationship can be harnessed as a decision-support tool for extended-range GL forecasts, based on purely statistical information and independent of NWP models. Additionally, our neural network approach shows particular value for predicting Blocking over Europe (EuBL and ScBL). Consistent with previous findings that NWP models struggle with these regimes -- often performing worse than simple climatological forecasts \citep[e.g.,][]{Bueler2021, Osman2023, Wandel2024} -- we find that two of our neural networks, one excluding and one including NWP information, both outperform the raw NWP forecasts for these blocked regimes. These results highlight that predictability sources should not be approached with a "one-size-fits-all" strategy but rather tailored to specific weather regimes. This perspective may have uncovered previously unknown sources of skill, such as the Antarctic Oscillation index’s potential relevance for European and Scandinavian Blocking.

Further research should focus on refining predictor selection for individual weather regimes to establish robust connections. One approach could be incorporating time-series analysis of indices in more complex machine learning models such as transformers to capture temporal dependencies. Additionally, conducting case studies on specific forecasts and their associated atmospheric states may provide deeper insights into the mechanisms driving predictability. Another important step is extending the analysis to longer lead times within the extended range to evaluate how different predictors contribute to forecast skill over increasing time horizons. By enhancing our understanding of regime-specific sources of predictability, we can further improve extended-range forecasting and develop more reliable decision-support tools for subseasonal-to-seasonal prediction.

To conclude, our study demonstrates the potential of statistical–dynamical forecasting with neural networks for the subseasonal prediction of North Atlantic-European weather regimes, providing valuable improvements to NWP performance -- particularly in situations where traditional NWP models struggle.

\section*{Conflict of interest}
All authors declare that they have no conflicts of interest.

\section*{Code and Data Availability}
Code is available on the GitHub repository \url{https://github.com/fmockert/windowsofopportunity_WR}. The ERA5 data can be obtained from the Climate Data Store \url{https://cds.climate.copernicus.eu/\#!/home}. Weather regime data are available from CMG upon request. The original S2S database is hosted at ECMWF as an extension of the TIGGE database.

\section*{Acknowledgements}\label{acknowledgements}
FM has received funding from the KIT Center for Mathematics in Sciences, Engineering and Economics under the seed funding programme. The contribution of JQ was partly funded by the European Union (ERC, ASPIRE, 101077260). The work of JQ and CMG was funded by the Helmholtz Association as part of the Young Investigator Group, “Sub-seasonal Predictability: Understanding the Role of Diabatic Outflow” (SPREADOUT, grant VH-NG-1243). SL gratefully acknowledges support by the Vector Stiftung through the Young Investigator Group “Artificial Intelligence for Probabilistic Weather Forecasting.”

\bibliography{bibliography/library_WoOpaper}

\begin{thebibliography}{45}
\providecommand{\natexlab}[1]{#1}
\providecommand{\url}[1]{\texttt{#1}}
\expandafter\ifx\csname urlstyle\endcsname\relax
  \providecommand{\doi}[1]{doi: #1}\else
  \providecommand{\doi}{doi: \begingroup \urlstyle{rm}\Url}\fi

\bibitem[Baldwin et~al.(2001)Baldwin, Gray, Dunkerton, Hamilton, Haynes, Randel, Holton, Alexander, Hirota, Horinouchi, Jones, Kinnersley, Marquardt, Sato, and Takahashi]{Baldwin2001}
M.~P. Baldwin, L.~J. Gray, T.~J. Dunkerton, K.~Hamilton, P.~H. Haynes, W.~J. Randel, J.~R. Holton, M.~J. Alexander, I.~Hirota, T.~Horinouchi, D.~B.~A. Jones, J.~S. Kinnersley, C.~Marquardt, K.~Sato, and M.~Takahashi.
\newblock The quasi-biennial oscillation.
\newblock 2001.
\newblock \doi{https://doi.org/10.1029/1999RG000073}.

\bibitem[Beerli and Grams(2019)]{Beerli2019}
R.~Beerli and C.~M. Grams.
\newblock Stratospheric modulation of the large-scale circulation in the atlantic–european region and its implications for surface weather events.
\newblock \emph{Quarterly Journal of the Royal Meteorological Society}, 145:\penalty0 3732--3750, 10 2019.
\newblock ISSN 1477870X.
\newblock \doi{10.1002/qj.3653}.

\bibitem[Bloomfield et~al.(2020)Bloomfield, Suitters, and Drew]{Bloomfield2020}
H.~C. Bloomfield, C.~C. Suitters, and D.~R. Drew.
\newblock Meteorological drivers of european power system stress.
\newblock \emph{Journal of Renewable Energy}, 2020:\penalty0 1--12, 8 2020.
\newblock ISSN 2314-4386.
\newblock \doi{10.1155/2020/5481010}.

\bibitem[Bloomfield et~al.(2021)Bloomfield, Brayshaw, Gonzalez, and Charlton-Perez]{Bloomfield2021b}
H.~C. Bloomfield, D.~J. Brayshaw, P.~L. Gonzalez, and A.~Charlton-Perez.
\newblock Pattern-based conditioning enhances sub-seasonal prediction skill of european national energy variables.
\newblock \emph{Meteorological Applications}, 28, 7 2021.
\newblock ISSN 14698080.
\newblock \doi{10.1002/met.2018}.

\bibitem[Büeler et~al.(2020)Büeler, Beerli, Wernli, and Grams]{Büeler2020}
D.~Büeler, R.~Beerli, H.~Wernli, and C.~M. Grams.
\newblock Stratospheric influence on ecmwf sub-seasonal forecast skill for energy-industry-relevant surface weather in european countries.
\newblock \emph{Quarterly Journal of the Royal Meteorological Society}, 146:\penalty0 3675--3694, 10 2020.
\newblock ISSN 1477870X.
\newblock \doi{10.1002/qj.3866}.

\bibitem[Büeler et~al.(2021)Büeler, Ferranti, Magnusson, Quinting, and Grams]{Bueler2021}
D.~Büeler, L.~Ferranti, L.~Magnusson, J.~F. Quinting, and C.~M. Grams.
\newblock Year-round sub-seasonal forecast skill for atlantic–european weather regimes.
\newblock \emph{Quarterly Journal of the Royal Meteorological Society}, 147:\penalty0 4283--4309, 10 2021.
\newblock ISSN 1477870X.
\newblock \doi{10.1002/qj.4178}.

\bibitem[Cassou(2008)]{Cassou2008}
C.~Cassou.
\newblock Intraseasonal interaction between the madden-julian oscillation and the north atlantic oscillation.
\newblock \emph{Nature}, 455:\penalty0 523--527, 9 2008.
\newblock ISSN 14764687.
\newblock \doi{10.1038/nature07286}.

\bibitem[Charlton-Perez et~al.(2019)Charlton-Perez, Aldridge, Grams, and Lee]{Charlton-Perez2019}
A.~J. Charlton-Perez, R.~W. Aldridge, C.~M. Grams, and R.~Lee.
\newblock Winter pressures on the uk health system dominated by the greenland blocking weather regime.
\newblock \emph{Weather and Climate Extremes}, 25, 9 2019.
\newblock ISSN 22120947.
\newblock \doi{10.1016/j.wace.2019.100218}.

\bibitem[Chase et~al.(2022)Chase, Harrison, Burke, Lackmann, and McGovern]{Chase2022}
R.~J. Chase, D.~R. Harrison, A.~Burke, G.~M. Lackmann, and A.~McGovern.
\newblock A machine learning tutorial for operational meteorology. part i: Traditional machine learning.
\newblock \emph{Weather and Forecasting}, 37:\penalty0 1509--1529, 8 2022.
\newblock ISSN 15200434.
\newblock \doi{10.1175/WAF-D-22-0070.1}.

\bibitem[Chase et~al.(2023)Chase, Harrison, Lackmann, and McGovern]{Chase2023}
R.~J. Chase, D.~R. Harrison, G.~M. Lackmann, and A.~McGovern.
\newblock A machine learning tutorial for operational meteorology. part ii: Neural networks and deep learning.
\newblock \emph{Weather and Forecasting}, 38:\penalty0 1271--1293, 8 2023.
\newblock ISSN 15200434.
\newblock \doi{10.1175/WAF-D-22-0187.1}.

\bibitem[Domeisen et~al.(2020)Domeisen, Grams, and Papritz]{Domeisen2020}
D.~I.~V. Domeisen, C.~M. Grams, and L.~Papritz.
\newblock The role of north atlantic–european weather regimes in the surface impact of sudden stratospheric warming events.
\newblock \emph{Weather and Climate Dynamics}, 1:\penalty0 373--388, 8 2020.
\newblock \doi{10.5194/wcd-1-373-2020}.

\bibitem[ECMWF(2018)]{ECMWF2018}
ECMWF.
\newblock Forecast ensemble (ens) - rationale and construction, 2018.
\newblock URL \url{https://confluence.ecmwf.int/display/FUG/Section+5+Forecast+Ensemble+%28ENS%29+-+Rationale+and+Construction}.

\bibitem[Elliot et~al.(2008)Elliot, Cross, and Fleming]{Elliot2008}
A.~J. Elliot, K.~W. Cross, and D.~M. Fleming.
\newblock Acute respiratory infections and winter pressures on hospital admissions in england and wales 1990-2005.
\newblock \emph{Journal of Public Health}, 30:\penalty0 91--98, 3 2008.
\newblock ISSN 17413842.
\newblock \doi{10.1093/pubmed/fdn003}.

\bibitem[Feng and Lin(2019)]{Feng2019}
P.~N. Feng and H.~Lin.
\newblock Modulation of the mjo-related teleconnections by the qbo.
\newblock \emph{Journal of Geophysical Research: Atmospheres}, 124:\penalty0 12022--12033, 11 2019.
\newblock ISSN 21698996.
\newblock \doi{10.1029/2019JD030878}.

\bibitem[Ferranti et~al.(2018)Ferranti, Magnusson, Vitart, and Richardson]{Ferranti2018}
L.~Ferranti, L.~Magnusson, F.~Vitart, and D.~S. Richardson.
\newblock How far in advance can we predict changes in large-scale flow leading to severe cold conditions over europe?
\newblock \emph{Quarterly Journal of the Royal Meteorological Society}, 144:\penalty0 1788--1802, 7 2018.
\newblock ISSN 1477870X.
\newblock \doi{10.1002/qj.3341}.

\bibitem[Gold et~al.(2020)Gold, White, Roeder, McAleenan, Kabban, and Ahner]{Gold2020}
S.~Gold, E.~White, W.~Roeder, M.~McAleenan, C.~S. Kabban, and D.~Ahner.
\newblock Probabilistic contingency tables: An improvement to verify probability forecasts.
\newblock \emph{Weather and Forecasting}, 35:\penalty0 609--621, 4 2020.
\newblock ISSN 15200434.
\newblock \doi{10.1175/WAF-D-19-0116.1}.

\bibitem[Grams et~al.(2017)Grams, Beerli, Pfenninger, Staffell, and Wernli]{Grams2017}
C.~M. Grams, R.~Beerli, S.~Pfenninger, I.~Staffell, and H.~Wernli.
\newblock Balancing europe's wind-power output through spatial deployment informed by weather regimes.
\newblock \emph{Nature Climate Change}, 7:\penalty0 557--562, 8 2017.
\newblock ISSN 17586798.
\newblock \doi{10.1038/NCLIMATE3338}.

\bibitem[Hauser et~al.(2023{\natexlab{a}})Hauser, Teubler, Riemer, Knippertz, and Grams]{Hauser2023}
S.~Hauser, F.~Teubler, M.~Riemer, P.~Knippertz, and C.~M. Grams.
\newblock Towards a holistic understanding of blocked regime dynamics through a combination of complementary diagnostic perspectives.
\newblock \emph{Weather and Climate Dynamics}, 4:\penalty0 399--425, 5 2023{\natexlab{a}}.
\newblock ISSN 26984016.
\newblock \doi{https://doi.org/10.5194/wcd-4-399-2023}.

\bibitem[Hauser et~al.(2023{\natexlab{b}})Hauser, Teubler, Riemer, Knippertz, and Grams]{Hauser2023pre}
S.~Hauser, F.~Teubler, M.~Riemer, P.~Knippertz, and C.~M. Grams.
\newblock Life cycle dynamics of greenland blocking from a potential vorticity perspective.
\newblock \emph{EGUsphere Preprint repository}, 2023{\natexlab{b}}.
\newblock \doi{https://doi.org/10.5194/egusphere-2023-2945}.
\newblock URL \url{https://doi.org/10.5194/egusphere-2023-2945}.

\bibitem[Hersbach et~al.(2020)Hersbach, Bell, Berrisford, Hirahara, Horányi, Muñoz-Sabater, Nicolas, Peubey, Radu, Schepers, Simmons, Soci, Abdalla, Abellan, Balsamo, Bechtold, Biavati, Bidlot, Bonavita, Chiara, Dahlgren, Dee, Diamantakis, Dragani, Flemming, Forbes, Fuentes, Geer, Haimberger, Healy, Hogan, Hólm, Janisková, Keeley, Laloyaux, Lopez, Lupu, Radnoti, d.~Rosnay, Rozum, Vamborg, Villaume, and Thépaut]{Hersbach2020}
H.~Hersbach, B.~Bell, P.~Berrisford, S.~Hirahara, A.~Horányi, J.~Muñoz-Sabater, J.~Nicolas, C.~Peubey, R.~Radu, D.~Schepers, A.~Simmons, C.~Soci, S.~Abdalla, X.~Abellan, G.~Balsamo, P.~Bechtold, G.~Biavati, J.~Bidlot, M.~Bonavita, G.~D. Chiara, P.~Dahlgren, D.~Dee, M.~Diamantakis, R.~Dragani, J.~Flemming, R.~Forbes, M.~Fuentes, A.~Geer, L.~Haimberger, S.~Healy, R.~J. Hogan, E.~Hólm, M.~Janisková, S.~Keeley, P.~Laloyaux, P.~Lopez, C.~Lupu, G.~Radnoti, P.~d.~Rosnay, I.~Rozum, F.~Vamborg, S.~Villaume, and J.~N. Thépaut.
\newblock The era5 global reanalysis.
\newblock \emph{Quarterly Journal of the Royal Meteorological Society}, 146:\penalty0 1999--2049, 7 2020.
\newblock ISSN 1477870X.
\newblock \doi{10.1002/qj.3803}.

\bibitem[Kent et~al.(2023)Kent, Scaife, and Dunstone]{Kent2022}
C.~Kent, A.~A. Scaife, and N.~Dunstone.
\newblock What potential for improving sub-seasonal predictions of the winter nao?
\newblock \emph{Atmospheric Science Letters}, 24, 4 2023.
\newblock ISSN 1530261X.
\newblock \doi{10.1002/asl.1146}.

\bibitem[Lee et~al.(2020)Lee, Lee, Woolnough, and Boxall]{Lee2020}
J.~C.~K. Lee, R.~W. Lee, S.~J. Woolnough, and L.~J. Boxall.
\newblock The links between the madden-julian oscillation and european weather regimes.
\newblock \emph{Theoretical and Applied Climatology}, 141:\penalty0 567--586, 7 2020.
\newblock ISSN 14344483.
\newblock \doi{10.1007/s00704-020-03223-2}.

\bibitem[Lee et~al.(2019)Lee, Woolnough, Charlton-Perez, and Vitart]{Lee2019}
R.~W. Lee, S.~J. Woolnough, A.~J. Charlton-Perez, and F.~Vitart.
\newblock Enso modulation of mjo teleconnections to the north atlantic and europe.
\newblock \emph{Geophysical Research Letters}, 46:\penalty0 13535--13545, 11 2019.
\newblock ISSN 19448007.
\newblock \doi{10.1029/2019GL084683}.

\bibitem[Lin et~al.(2009)Lin, Brunet, and Derome]{Lin2009}
H.~Lin, G.~Brunet, and J.~Derome.
\newblock An observed connection between the north atlantic oscillation and the madden-julian oscillation.
\newblock \emph{Journal of Climate}, 22:\penalty0 364--380, 1 2009.
\newblock ISSN 08948755.
\newblock \doi{10.1175/2008JCLI2515.1}.

\bibitem[Madden and Julian(1971)]{Madden1971}
R.~A. Madden and P.~R. Julian.
\newblock Detection of a 40–50 day oscillation in the zonal wind in the tropical pacific.
\newblock \emph{Journal of the Atmoshperic Sciences}, 28:\penalty0 702--708, 3 1971.
\newblock \doi{https://doi.org/10.1175/1520-0469(1971)028%3C0702:DOADOI%3E2.0.CO;2}.

\bibitem[Michel and Rivière(2011)]{Michel2011}
C.~Michel and G.~Rivière.
\newblock The link between rossby wave breakings and weather regime transitions.
\newblock \emph{Journal of the Atmospheric Sciences}, 68:\penalty0 1730--1748, 8 2011.
\newblock ISSN 00224928.
\newblock \doi{10.1175/2011JAS3635.1}.

\bibitem[Michelangeli et~al.(1995)Michelangeli, Vautard, and Legras]{Michelangeli1995}
P.~Michelangeli, R.~Vautard, and B.~Legras.
\newblock Weather regimes: Recurrence and quasi stationarity.
\newblock \emph{Journal of Atmospheric Sciences}, 52:\penalty0 1237--1256, 1995.
\newblock \doi{https://doi.org/10.1175/1520-0469(1995)052%3C1237:WRRAQS%3E2.0.CO;2}.

\bibitem[Millin et~al.(2024)Millin, Furtado, and Malloy]{Millin2024}
O.~T. Millin, J.~C. Furtado, and C.~Malloy.
\newblock The impact of north american winter weather regimes on electricity load in the central united states.
\newblock \emph{npj Climate and Atmospheric Science}, 7, 12 2024.
\newblock ISSN 23973722.
\newblock \doi{10.1038/s41612-024-00803-1}.

\bibitem[Mockert et~al.(2023)Mockert, Grams, Brown, and Neumann]{Mockert2023}
F.~Mockert, C.~M. Grams, T.~Brown, and F.~Neumann.
\newblock Meteorological conditions during periods of low wind speed and insolation in germany: The role of weather regimes.
\newblock \emph{Meteorological Applications}, 30, 7 2023.
\newblock ISSN 1350-4827.
\newblock \doi{10.1002/met.2141}.
\newblock URL \url{https://rmets.onlinelibrary.wiley.com/doi/10.1002/met.2141}.

\bibitem[Mockert et~al.(2024)Mockert, Grams, Lerch, Osman, and Quinting]{Mockert2024}
F.~Mockert, C.~M. Grams, S.~Lerch, M.~Osman, and J.~Quinting.
\newblock Multivariate post‐processing of probabilistic sub‐seasonal weather regime forecasts.
\newblock \emph{Quarterly Journal of the Royal Meteorological Society}, 9 2024.
\newblock ISSN 0035-9009.
\newblock \doi{10.1002/qj.4840}.
\newblock URL \url{https://rmets.onlinelibrary.wiley.com/doi/10.1002/qj.4840}.

\bibitem[NOAA(2025)]{NOAA2025b}
NOAA.
\newblock Daily mjo index time series, 2 2025.
\newblock URL \url{https://psl.noaa.gov/mjo/mjoindex/}.

\bibitem[Osman et~al.(2023)Osman, Beerli, Büeler, and Grams]{Osman2023}
M.~Osman, R.~Beerli, D.~Büeler, and C.~M. Grams.
\newblock Multi‐model assessment of sub‐seasonal predictive skill for year‐round atlantic‐european weather regimes.
\newblock \emph{Quarterly Journal of the Royal Meteorological Society}, 7 2023.
\newblock ISSN 0035-9009.
\newblock \doi{https://doi.org/10.1002/qj.4512}.

\bibitem[Roberts et~al.(2023)Roberts, Balmaseda, Ferranti, and Vitart]{Roberts2023}
C.~D. Roberts, M.~A. Balmaseda, L.~Ferranti, and F.~Vitart.
\newblock Euro-atlantic weather regimes and their modulation by tropospheric and stratospheric teleconnection pathways in ecmwf reforecasts.
\newblock \emph{Monthly Weather Review}, 151:\penalty0 2779--2799, 2023.
\newblock \doi{10.1175/MWR-D}.
\newblock URL \url{https://doi.org/10.1175/MWR-D-}.

\bibitem[Salient et~al.(2023)Salient, Deo, and Atlas]{Salient2023}
Salient, M.~Deo, and A.~Atlas.
\newblock On the money: How salient's novel s2s forecasts can save the agriculture industry millions of dollars, 5 2023.
\newblock URL \url{https://www.salientpredictions.com/blog/on-the-money-how-salients-novel-s2s-forecasts-can-save-the-agriculture-industry-millions-of-dollars}.

\bibitem[Scaife et~al.(2022)Scaife, Hermanson, v.~Niekerk, Andrews, Baldwin, Belcher, Bett, Comer, Dunstone, Geen, Hardiman, Ineson, Knight, Nie, Ren, and Smith]{Scaife2022}
A.~A. Scaife, L.~Hermanson, A.~v.~Niekerk, M.~Andrews, M.~P. Baldwin, S.~Belcher, P.~Bett, R.~E. Comer, N.~J. Dunstone, R.~Geen, S.~C. Hardiman, S.~Ineson, J.~Knight, Y.~Nie, H.~L. Ren, and D.~Smith.
\newblock Long-range predictability of extratropical climate and the length of day.
\newblock \emph{Nature Geoscience}, 10 2022.
\newblock ISSN 17520908.
\newblock \doi{10.1038/s41561-022-01037-7}.

\bibitem[Song et~al.(2009)Song, Zhou, Li, and Qi]{Song2009}
J.~Song, W.~Zhou, C.~Li, and L.~Qi.
\newblock Signature of the antarctic oscillation in the northern hemisphere.
\newblock \emph{Meteorology and Atmospheric Physics}, 105:\penalty0 55--67, 2009.
\newblock ISSN 01777971.
\newblock \doi{10.1007/s00703-009-0036-5}.

\bibitem[Spaeth et~al.(2024)Spaeth, Rupp, Osman, Grams, and Birner]{Spaeth2024}
J.~Spaeth, P.~Rupp, M.~Osman, C.~M. Grams, and T.~Birner.
\newblock Flow‐dependence of ensemble spread of subseasonal forecasts explored via north atlantic‐european weather regimes.
\newblock \emph{Geophysical Research Letters}, 51, 7 2024.
\newblock ISSN 0094-8276.
\newblock \doi{10.1029/2024GL109733}.
\newblock URL \url{https://agupubs.onlinelibrary.wiley.com/doi/10.1029/2024GL109733}.

\bibitem[Specq and Batté(2022)]{Specq2022}
D.~Specq and L.~Batté.
\newblock Do subseasonal forecasts take advantage of madden–julian oscillation windows of opportunity?
\newblock \emph{Atmospheric Science Letters}, 23, 4 2022.
\newblock ISSN 1530261X.
\newblock \doi{10.1002/asl.1078}.

\bibitem[Tachibana et~al.(2018)Tachibana, Inoue, Komatsu, Nakamura, Honda, Ogata, and Yamazaki]{Tachibana2018}
Y.~Tachibana, Y.~Inoue, K.~K. Komatsu, T.~Nakamura, M.~Honda, K.~Ogata, and K.~Yamazaki.
\newblock Interhemispheric synchronization between the ao and the aao.
\newblock \emph{Geophysical Research Letters}, 45:\penalty0 13,477--13,484, 12 2018.
\newblock ISSN 19448007.
\newblock \doi{10.1029/2018GL081002}.

\bibitem[Vautard(1990)]{Vautard1990}
R.~Vautard.
\newblock Multiple weather regimes over the north atlantic: Analysis of precursors and successors.
\newblock \emph{Monthly Weather Review}, 118:\penalty0 2056--2081, 1990.
\newblock \doi{https://doi.org/10.1175/1520-0493(1990)118%3C2056:MWROTN%3E2.0.CO;2}.

\bibitem[Vitart(2017)]{Vitart2017b}
F.~Vitart.
\newblock Madden—julian oscillation prediction and teleconnections in the s2s database.
\newblock \emph{Quarterly Journal of the Royal Meteorological Society}, 143:\penalty0 2210--2220, 7 2017.
\newblock ISSN 1477870X.
\newblock \doi{10.1002/qj.3079}.

\bibitem[Vitart et~al.(2017)Vitart, Ardilouze, Bonet, Brookshaw, Chen, Codorean, Déqué, Ferranti, Fucile, Fuentes, Hendon, Hodgson, Kang, Kumar, Lin, Liu, Liu, Malguzzi, Mallas, Manoussakis, Mastrangelo, MacLachlan, McLean, Minami, Mladek, Nakazawa, Najm, Nie, Rixen, Robertson, Ruti, Sun, Takaya, Tolstykh, Venuti, Waliser, Woolnough, Wu, Won, Xiao, Zaripov, and Zhang]{Vitart2017}
F.~Vitart, C.~Ardilouze, A.~Bonet, A.~Brookshaw, M.~Chen, C.~Codorean, M.~Déqué, L.~Ferranti, E.~Fucile, M.~Fuentes, H.~Hendon, J.~Hodgson, H.~S. Kang, A.~Kumar, H.~Lin, G.~Liu, X.~Liu, P.~Malguzzi, I.~Mallas, M.~Manoussakis, D.~Mastrangelo, C.~MacLachlan, P.~McLean, A.~Minami, R.~Mladek, T.~Nakazawa, S.~Najm, Y.~Nie, M.~Rixen, A.~W. Robertson, P.~Ruti, C.~Sun, Y.~Takaya, M.~Tolstykh, F.~Venuti, D.~Waliser, S.~Woolnough, T.~Wu, D.~J. Won, H.~Xiao, R.~Zaripov, and L.~Zhang.
\newblock The subseasonal to seasonal (s2s) prediction project database.
\newblock \emph{Bulletin of the American Meteorological Society}, 98:\penalty0 163--173, 1 2017.
\newblock ISSN 00030007.
\newblock \doi{10.1175/BAMS-D-16-0017.1}.

\bibitem[Wandel et~al.(2024)Wandel, Büeler, Knippertz, Quinting, and Grams]{Wandel2024}
J.~Wandel, D.~Büeler, P.~Knippertz, J.~F. Quinting, and C.~M. Grams.
\newblock Why moist dynamic processes matter for the sub-seasonal prediction of atmospheric blocking over europe.
\newblock \emph{Journal of Geophysical Research: Atmospheres}, 129, 4 2024.
\newblock ISSN 21698996.
\newblock \doi{10.1029/2023JD039791}.

\bibitem[Wheeler and Hendon(2004)]{Wheeler2004}
M.~C. Wheeler and H.~H. Hendon.
\newblock An all-season real-time multivariate mjo index: Development of an index for monitoring and prediction.
\newblock 2004.
\newblock \doi{https://doi.org/10.1175/1520-0493(2004)132%3C1917:AARMMI%3E2.0.CO;2}.

\bibitem[White et~al.(2017)White, Carlsen, Robertson, Klein, Lazo, Kumar, Vitart, d.~Perez, Ray, Murray, Bharwani, MacLeod, James, Fleming, Morse, Eggen, Graham, Kjellström, Becker, Pegion, Holbrook, McEvoy, Depledge, Perkins-Kirkpatrick, Brown, Street, Jones, Remenyi, Hodgson-Johnston, Buontempo, Lamb, Meinke, Arheimer, and Zebiak]{White2017}
C.~J. White, H.~Carlsen, A.~W. Robertson, R.~J. Klein, J.~K. Lazo, A.~Kumar, F.~Vitart, E.~C. d.~Perez, A.~J. Ray, V.~Murray, S.~Bharwani, D.~MacLeod, R.~James, L.~Fleming, A.~P. Morse, B.~Eggen, R.~Graham, E.~Kjellström, E.~Becker, K.~V. Pegion, N.~J. Holbrook, D.~McEvoy, M.~Depledge, S.~Perkins-Kirkpatrick, T.~J. Brown, R.~Street, L.~Jones, T.~A. Remenyi, I.~Hodgson-Johnston, C.~Buontempo, R.~Lamb, H.~Meinke, B.~Arheimer, and S.~E. Zebiak.
\newblock Potential applications of subseasonal-to-seasonal (s2s) predictions.
\newblock \emph{Meteorological Applications}, 24:\penalty0 315--325, 7 2017.
\newblock ISSN 14698080.
\newblock \doi{https://doi.org/10.1002/met.1654}.

\end{thebibliography}
\newpage

\renewcommand{\thesection}{S}
\section{Supplementary material}
\setcounter{figure}{0}
\renewcommand\thefigure{\thesection\arabic{figure}}

\setcounter{table}{0}
\renewcommand\thetable{\thesection\arabic{table}}

\begin{figure}[!h]   
\centering
\begin{subfigure}[c]{0.9\textwidth}
\subcaption{MJO-conditioned rates.}
\includegraphics[width=1\linewidth]{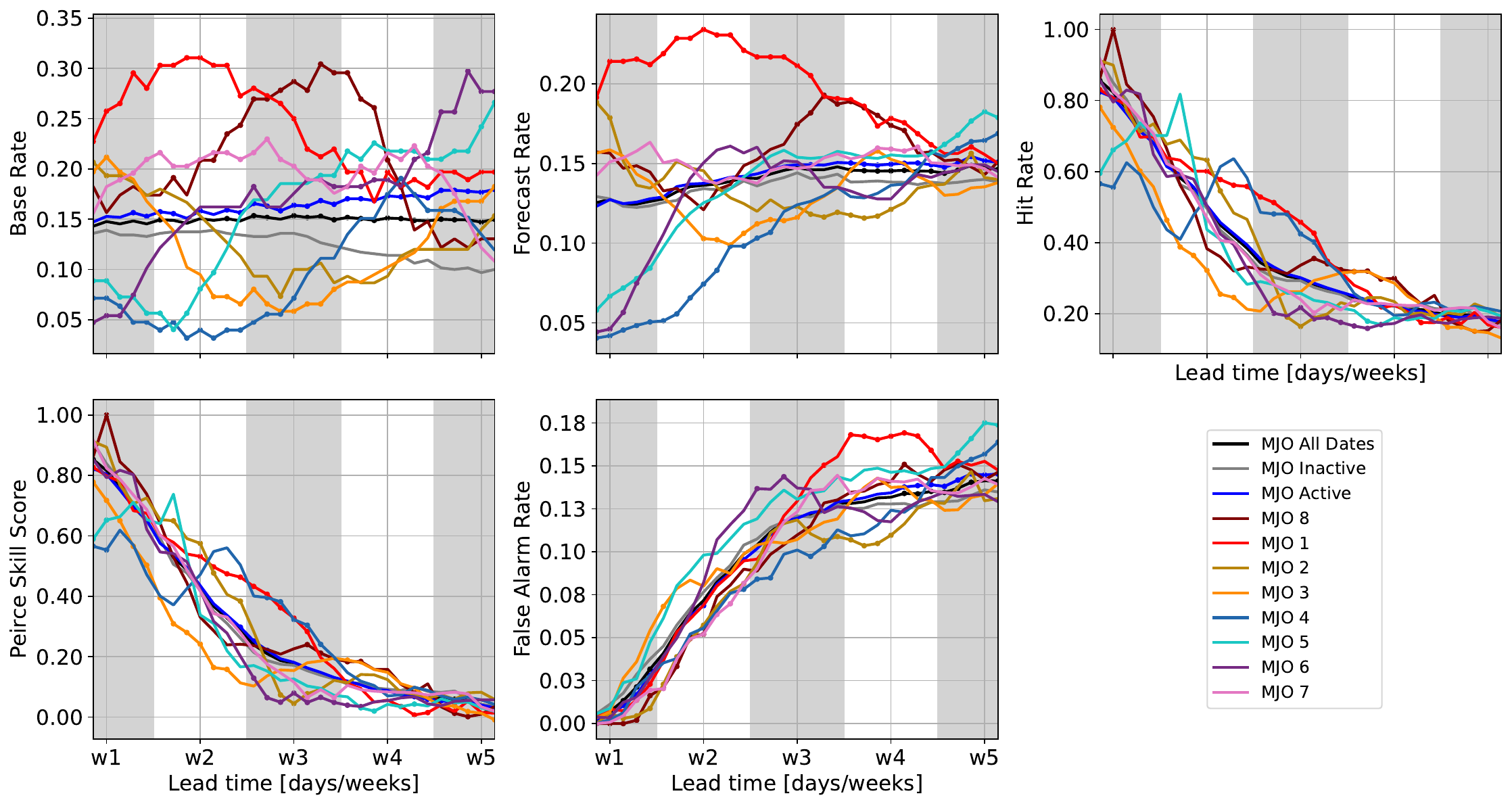}
\end{subfigure}\quad

\begin{subfigure}[c]{0.9\textwidth}
\subcaption{SPV-conditioned rates.}
\includegraphics[width=1\linewidth]{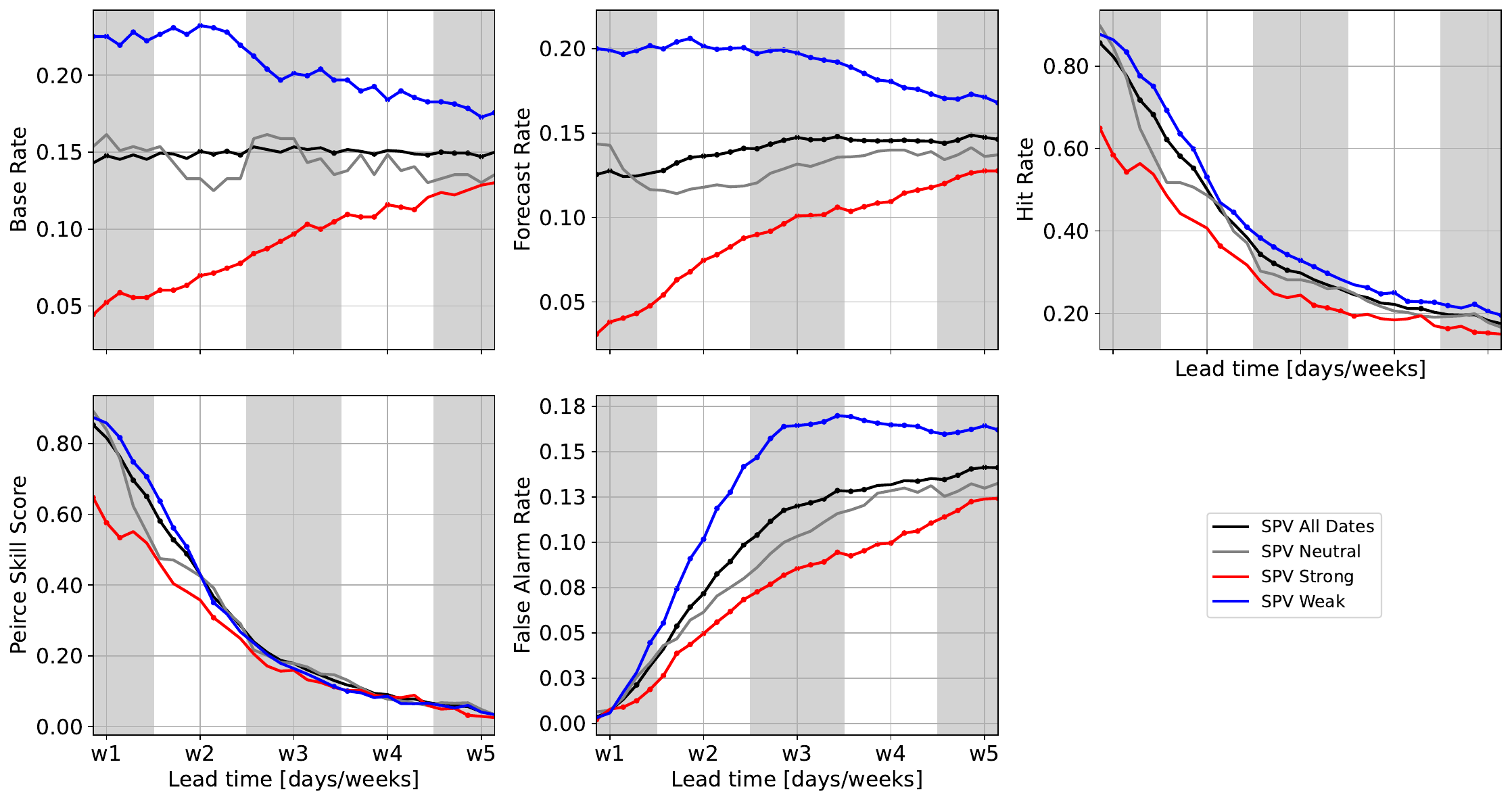}
\end{subfigure}\quad
\caption{Rates for weekly mean Greenland Blocking activity ($WRact_{mean, GL}$) in the extended winter period (November--March). These rates are split into subsets based on phases of the (a) MJO or (b) SPV. In contrast to Figure \ref{fig:baserates_GLactivity} and Figure \ref{fig:WoO_GLactivity}, here all lead times are shown (x-axis). Dots on the individual lines indicate whether the rates in the specific categories and at the specific lead time are significantly different to their inactive/neutral phases -- for MJO and SPV, respectively -- with a 90\% confidence interval.
}
\label{fig:rates_GLactivity_allleads}
\end{figure}

\begin{figure}[!h]
    \centering
    \includegraphics[width=1\linewidth]{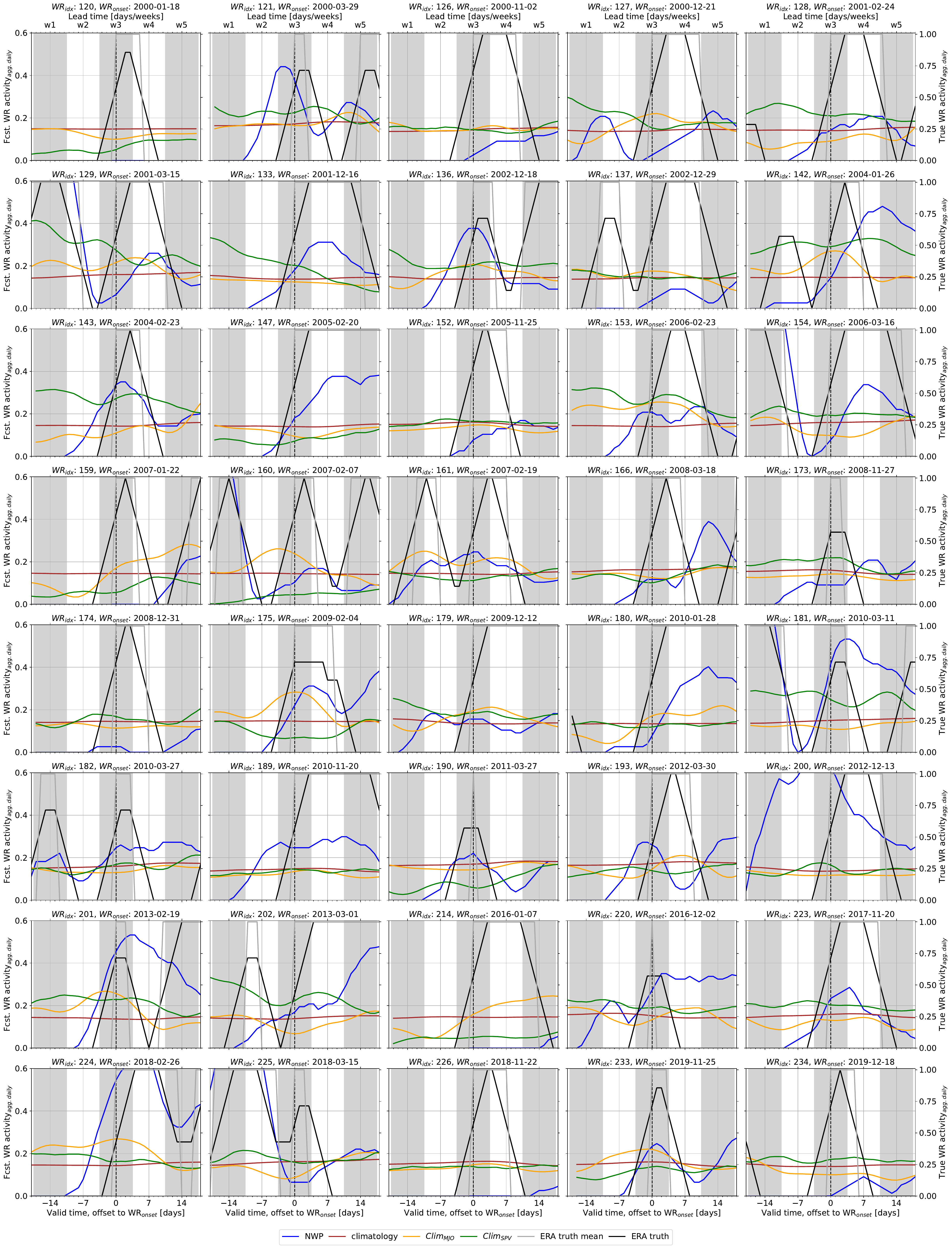}
    \caption{Forecasts of the aggregated daily Greenland Blocking activity used for the mean composite plot in Figure \ref{fig:AroundWRonset}a for all Greenland Blocking activities. The forecasts are initialised three weeks ahead of the Greenland Blocking activity onset.
    }
    \label{fig:AroundWRonset_individual_singlefcst}
\end{figure}

\begin{figure}[!h]
    \centering
    \includegraphics[width=1\linewidth]{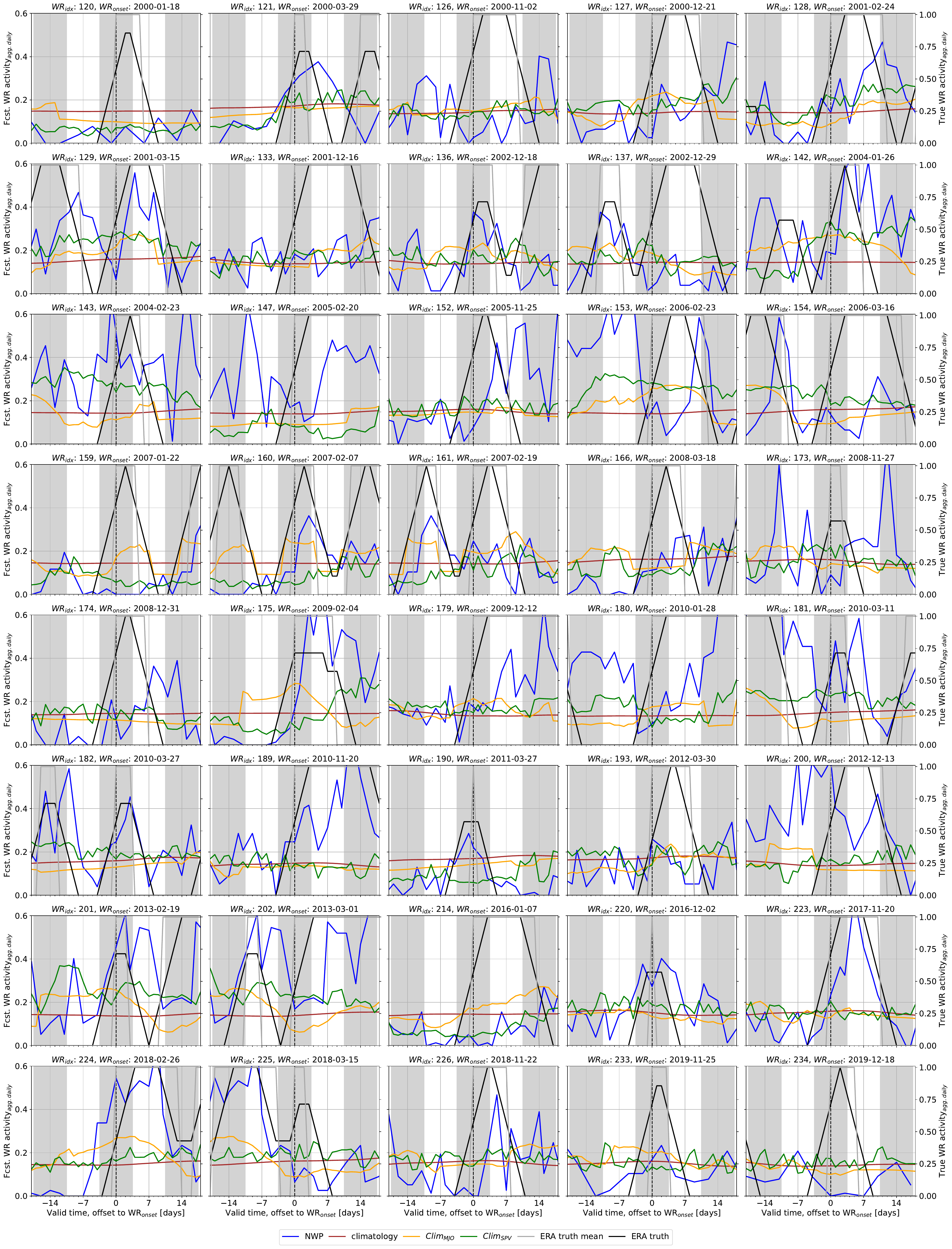}
    \caption{Forecasts of the aggregated daily Greenland Blocking activity used for the mean composite plot in Figure \ref{fig:AroundWRonset}b for all Greenland Blocking activities. Consecutive forecasts with each a lead time of three weeks are used.
    }
    \label{fig:AroundWRonset_individual_multifcst}
\end{figure}

\begin{figure}[!h]
    \centering
    \includegraphics[width=1.0\linewidth]{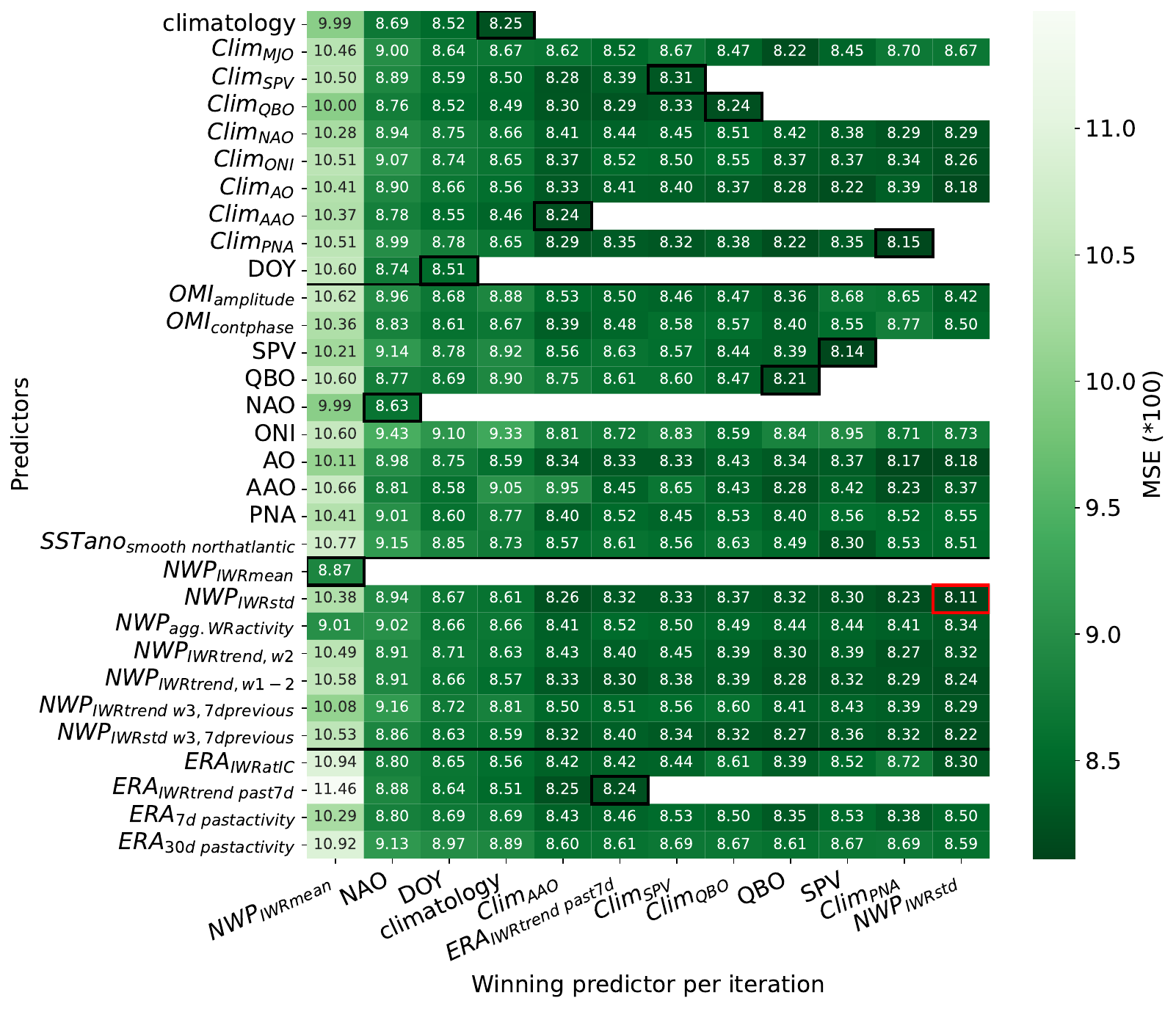}
    \caption{Visualisation of the stepwise feature selection process for the all-features neural network ($NN_{all}$) model and Greenland Blocking activity for forecast week three. The x-axis indicates the winning feature per step of the feature selection. The selection process is defined via the MSE (visualised by colour and values, where the values are multiplied by 100 for better readability). The winning feature (best MSE score) is indicated for each step with black boxes and the overall winning combination is indicated with a red box. A detailed explanation of the features can be found in Table \ref{tab:featurelist1} and \ref{tab:featurelist2}.
    }
    \label{fig:featureselection_GL_ABCNWP}
\end{figure}

\begin{figure}[!h]
    \centering
    \includegraphics[width=1.0\linewidth]{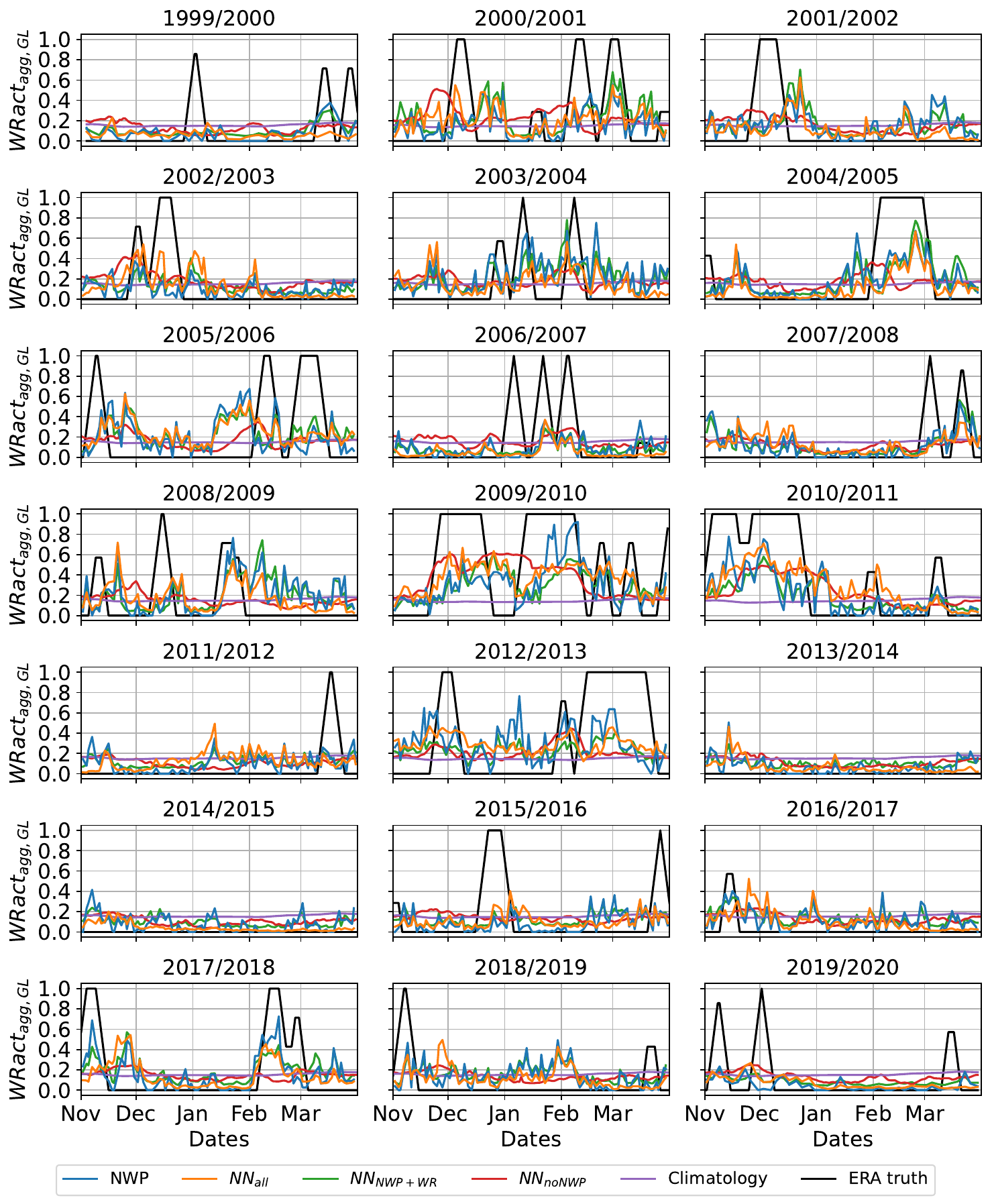}
    \caption{Forecasts of the aggregated daily GL activity for each extended winter period from 1999 to 2020. The ERA5 (actual) weather regime activity is indicated in black, the neural networks all-features, NWP and ERA, and climatology in blue, orange and green, as well as the NWP model in red and a climatological forecast in purple.
    }
    \label{fig:forecast_timeline}
\end{figure}

\begin{table}[ht]
\centering
\begin{tabular}{|l|l|p{8.5cm}|}  
\hline
\textbf{Category} & \textbf{Feature} & \textbf{Description} \\ \hline
\multirow{1}{*}{Climatology} 
    & Climatology   & Standard 91-days climatology \\ \cline{2-3} 
    & Clim\textsubscript{MJO}   & MJO-conditioned climatological forecast using OMI1/2 \\ \cline{2-3} 
    & Clim\textsubscript{SPV}   & SPV-conditioned climatological forecast using 100 hPa zonal mean wind at 60°N \\ \cline{2-3} 
    & Clim\textsubscript{QBO}   & QBO-conditioned climatological forecast using Quasi-Biennial Oscillation index \\ \cline{2-3} 
    & Clim\textsubscript{NAO}   & NAO-conditioned climatological forecast using North Atlantic Oscillation index \\ \cline{2-3} 
    & Clim\textsubscript{ONI}   & ONI-conditioned climatological forecast using Ocean Nino index \\ \cline{2-3} 
    & Clim\textsubscript{AO}    & AO-conditioned climatological forecast using Arctic Oscillation index \\ \cline{2-3} 
    & Clim\textsubscript{AAO}   & AAO-conditioned climatological forecast using Antarctic Oscillation index \\ \cline{2-3} 
    & Clim\textsubscript{PNA}   & PNA-conditioned climatological forecast using Pacific North-American index \\ \cline{2-3} 
    & DOY                     & Day of year represented by sinus curve (maxima at July 15th and minima at January 15th) \\ \hline
    
\multirow{1}{*}{Atmospheric State}
    & OMI\textsubscript{amplitude}  & MJO amplitude in OMI-phase space. Source: \url{https://psl.noaa.gov/mjo/mjoindex/omi.era5.1x.webpage.4023.txt} \\ \cline{2-3} 
    & OMI\textsubscript{contphase}  & MJO angle in OMI-phase space. Source: \url{} \\ \cline{2-3} 
    & QBO                     & QBO index. Source: \url{https://www.geo.fu-berlin.de/met/ag/strat/produkte/qbo/qbo.dat} \\ \cline{2-3} 
    & ONI                     & ONI index. Source: \url{https://psl.noaa.gov/data/correlation/oni.data} \\ \cline{2-3} 
    & NAO                     & NAO index. Source: \url{https://www.cpc.ncep.noaa.gov/products/precip/CWlink/pna/norm.nao.monthly.b5001.current.ascii.table} \\ \cline{2-3} 
    & AO                      & AO index. Source: \url{https://ftp.cpc.ncep.noaa.gov/cwlinks/norm.daily.ao.index.b500101.current.ascii} \\ \cline{2-3} 
    & PNA                     & PNA index. Source: \url{https://ftp.cpc.ncep.noaa.gov/cwlinks/norm.daily.pna.index.b500101.current.ascii} \\ \cline{2-3} 
    & AAO                     & AAO index. Source: \url{https://ftp.cpc.ncep.noaa.gov/cwlinks/norm.daily.aao.index.b790101.current.ascii} \\ \cline{2-3} 
    & SPV                     & SPV index computed by 60°N zonal mean wind at 100 hPa \\ \cline{2-3} 
    & SSTano\textsubscript{smooth northatlantic} & 30-day running mean anomalies of seas surface temperature compared to 1991--2020 in the North Atlantic (0 to 80°N, 80 to 10°W) \\ \hline
\end{tabular}
\caption{Overview of features available for neural network selection, categorised into Climatology and Atmospheric State. Features from the NWP and recent weather regime activity categories are listed separately in Table \ref{tab:featurelist2}.}
\label{tab:featurelist1}
\end{table}

\begin{table}[ht]
\centering
\begin{tabular}{|l|l|p{8.5cm}|}  
\hline
\textbf{Category} & \textbf{Feature} & \textbf{Description} \\ \hline
\multirow{1}{*}{NWP} 
    & NWP\textsubscript{IWRmean}       & Mean weather regime index forecast for week 3 \\ \cline{2-3} 
    & NWP\textsubscript{IWRstd}        & Standard deviation of ensemble weather regime index forecast for week 3 \\ \cline{2-3} 
    & NWP\textsubscript{agg. WRactivity} & Aggregated daily weather regime activity forecast for week 3 \\ \cline{2-3} 
    & NWP\textsubscript{IWRtrend,w2}    & Trend of daily IWR forecast across week 2 (linear regression) \\ \cline{2-3} 
    & NWP\textsubscript{IWRtrend,w1-2}  & Trend of daily IWR forecast across week 1-2 (linear regression) \\ \cline{2-3} 
    & NWP\textsubscript{IWRtrend,w3,7dprevious} & Trend of the weather regime index across the forecasts available 7d prior to initialisation time and a valid time similar to the 3 week forecast from initialisation time \\ \cline{2-3} 
    & NWP\textsubscript{IWRstd,w3,7dprevious} & Standard deviation of the weather regime index across the forecasts available 7d prior to initialisation time and a valid time similar to the 3 week forecast from initialisation time \\ \hline
    
\multirow{1}{*}{Recent WR}
    & ERA\textsubscript{IWRatIC}         & IWR at initialisation time \\ \cline{2-3} 
    & ERA\textsubscript{IWRtrend,past7d}  & IWR trend (linear regression) of past 7 days \\ \cline{2-3} 
    & ERA\textsubscript{7d,pastactivity} & Mean aggregated daily weather regime activity of past 7 days \\ \cline{2-3} 
    & ERA\textsubscript{30d,pastactivity}& Mean aggregated daily weather regime activity of past 30 days \\ \hline
\end{tabular}
\caption{Overview of features available for neural network selection, categorised into NWP and recent weather regime activity. Features from the Climatology and Atmospheric State categories are listed separately in Table \ref{tab:featurelist1}.}
\label{tab:featurelist2}
\end{table}

\end{document}